\documentclass[a4paper,fleqn,12pt]{wlscirep}
\usepackage[utf8]{inputenc}
\usepackage[T1]{fontenc}
\usepackage[english]{babel}
\usepackage{amsmath}
\usepackage{amsfonts}
\usepackage{mathtools}
\usepackage{graphicx}
\usepackage{hyperref}

\newcommand{\bl}{\color{blue}}

\newcommand{\supplementarysection}{%
  \setcounter{figure}{0}
  \let\oldthefigure\thefigure
  \renewcommand{\thefigure}{S\oldthefigure}
  \section{Supplementary section}
  \let\oldchapter\chapter
  \renewcommand{\chapter}{
    \let\thefigure\oldthefigure
    \let\chapter\oldchapter
    \oldchapter
  }
}

\title{Systemic risk approach to mitigate delay cascading in railway networks}

\author[ae]{Simone Daniotti}
\author[a]{Vito D.~P.~Servedio}
\author[b]{Johannes Kager}
\author[b]{Aad Robben-Baldauf}
\author[acd*]{Stefan Thurner}
\affil[a]{Complexity Science Hub Vienna, Josefst\"adter Strasse 39, 1080 Vienna, Austria}
\affil[b]{\"OBB-Personenverkehr AG, Am Hauptbahnhof 2, 1100 Vienna}
\affil[c]{Section for Science of Complex Systems, CeMSIIS, Medical University of Vienna, Spitalgasse 23, 1090, Austria}
\affil[d]{Santa Fe Institute, 1399 Hyde Park Road, Santa Fe, NM 85701, United States}
\affil[e]{Vienna University of Technology, Informatics, Vienna, 1040, Austria}
\affil[*]{ {stefan.thurner@meduniwien.ac.at}}

\begin{abstract}
In public railway systems, minor disruptions can trigger cascading events that lead to delays in the entire system. Typically, delays originate and propagate because the equipment is blocking ways, operational units are unavailable, or at the wrong place at the needed time. The specific understanding of the origins and processes involved in delay-spreading is still a challenge, even though large-scale simulations of national railway systems are becoming available on a highly detailed scale. Without this understanding, efficient management of delay propagation, a growing concern in some Western countries, will remain impossible. Here, we present a systemic risk-based approach to manage daily delay cascading on national scales. We compute the {\em systemic impact} of every train as the maximum of all delays it could possibly cause due to its interactions with other trains, infrastructure, and operational units. To compute it, we design an effective impact network where nodes are train services and links represent interactions that could cause delays. Our results are not only consistent with highly detailed and computationally intensive agent-based railway simulations but also allow us to pinpoint and identify the causes of delay cascades in detail. The systemic approach reveals structural weaknesses in railway systems whenever shared resources are involved. We use the systemic impact to optimally allocate additional shared resources to the system to reduce delays with minimal costs and effort. 
The method offers a practical and intuitive solution for delay management by optimizing the effective impact network through the introduction of new cheap local train services.
\end{abstract}

\begin{document}

\maketitle

\noindent
The functioning of modern societies counts on the reliable performance of various socio-technological systems, such as trade networks \cite{de2011world}, financial networks \cite{boss2004network, thurner2013debtrank}, supply chains~\cite{Harland2001ATO}, energy production~\cite{glassman2011geo}, and human mobility~\cite{alessandretti2020scales}. Failures in these networked systems usually come at tremendous social and monetary costs since their consequences may affect the entire system: they are systemic~\cite{battiston2016price,al2016economic}. An example is the 2003 Northeast Blackout in the United States, where the interconnected power grid failure affected 55 million people across eight states and cost an estimated 6 billion dollars. In the context of transportation, the Association of American Railroads estimated that rail service interruptions in transportation systems could cost the US economy up to 2 Billion dollars per day. Large-scale rail disruptions can lead to severe economic damage by disrupting cargo and passenger transport \cite{tsuchiya2007economic}, especially in densely urbanized countries where railway transport has developed into an inextricable asset for societal and economic well-being. 

System-wide failure in networked systems is often associated with cascading failure processes~\cite{boss2004contagion,Hirshleifer2003HerdBA,Crucitti2003ModelFC,BorgeHolthoefer2013CascadingBI}. The chance for those to appear as well as their expected size are what has become as known as {\em systemic risk}. Systemic risk and their corresponding failure distributions are often characterized by power-law statistics~\cite{thurner2018introduction} that often result from a networked structure of the underlying system.
Fat-tailed delay distributions in public transport networks are one example and have been empirically observed for a long time~\cite{briggs2007modelling,monechiComplexDelayDynamics2018}. 
Networks of networks add another aspect of systemic risk. 
For example, the study of the coupling between the European power grid and the Internet grid~\cite{havlin2014vulnerability} shows that the shutdown of only\ 10\% of the power stations and a cut of\ 12\% of server providers would cause\ 90\% of nodes to fail. Generally, the coupling of networks leads to discontinuous transitions that manifest themselves in massive failures with minimal chances to predict them.  

The causes for delays in public transport systems such as national railways can be categorized into \emph{exogenous} or \emph{primary delays}, driven by external factors, and \emph{endogenous} or \emph{secondary} delays that are created by the interactions between agents in the system~\cite{monechiComplexDelayDynamics2018}, usually through the scarcity of available resources~\cite{dekker2021cascading}.
The primary interest in most railway studies has been on identifying risks associated with disruptions happening at the infrastructure level~\cite{bhatia2015network}. In Europe, the efficiency of strategies mostly depends on how countries organize personnel and train in disruptive situations~\cite{schipperDifferencesSimilaritiesEuropean2018}.

Over the past years, modeling and simulating such systems brought new insights into the robustness of interconnected networks in terms of cascades: in~\cite{buldyrev2010catastrophic}, an analytical framework to study cascades of failure in interdependent networks has been developed.
In railways, a theoretical model of interactions between the personnel and the rolling stock layer~\cite{ball2016two} has been able to accurately display transitions in the system and predict the collapse of service.
These studies show the relevance of network interconnectivity.
The interdependency between personnel and rolling stock to understand cascading effects during significant disruptions was explored in \cite{dekker2021cascading}, 
emphasizing the necessity of respective information when studying delay cascading.

New notions of transportation resilience \cite{leobons2019assessing} and perturbations in the network topology \cite{lordan2015robustness} have been introduced from a network perspective. It allows one to deal with disruptions in a new light including network centrality measures. 
Topological measures of the railway network (network of stations), were applied to quantify the resilience and how prone train stations are to disruptions \cite{pagani2019resilience}. Studies on centrality measures in the context of freight transportation \cite{sun2019exploring} and on vulnerability using passenger destinations~\cite{wang2013vulnerability} highlighted the importance of the different agents acting on the networks.

Predicting delays and disruptions based on one train being the cause for the delay of another is a central problem. Bayesian networks were used for predicting delays~\cite{kecman2015stochastic}, showing how statistics and the interconnection and influence of the different agents acting on the railway infrastructure can accurately predict the delays happening on a normal working day.
Also, models explicitly implementing mechanisms of delay propagation in the network of stations have been proposed to study and predict the evolution of delays in a disruptive situation, like diffusion~\cite{dekker2022modelling} or epidemic spreading models~\cite{monechiComplexDelayDynamics2018}.
Moreover, these kinds of systems lend themselves to applications in machine learning or big-data approaches \cite{dekkerPredictingTransitionsMacroscopic2019,onetoTrainDelayPrediction2018} for predictions on their future status and to spot disruptive transitions in advance. 
Railways are systems in which a single operation (a train trajectory), which depends on the coordinated presence of different events at the same time (personnel and rolling stock available, free infrastructure), may lead to a sequence of concatenated disruptions~\cite{fleurquin2013systemic,ludvigsen2014extreme}.

\begin{figure}[th!]
\centering
\includegraphics[width=0.45\linewidth]{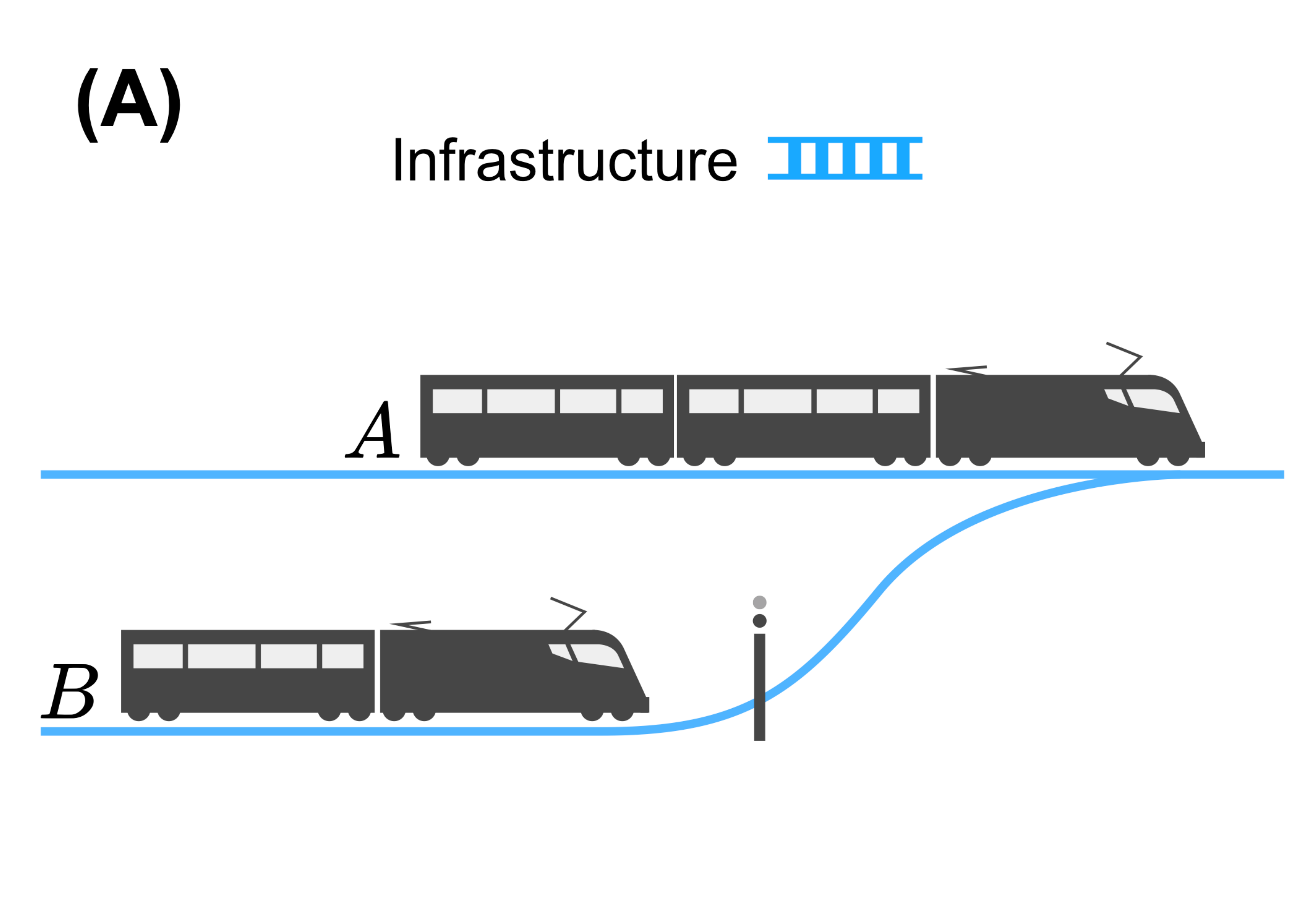}
\includegraphics[width=0.45\linewidth]{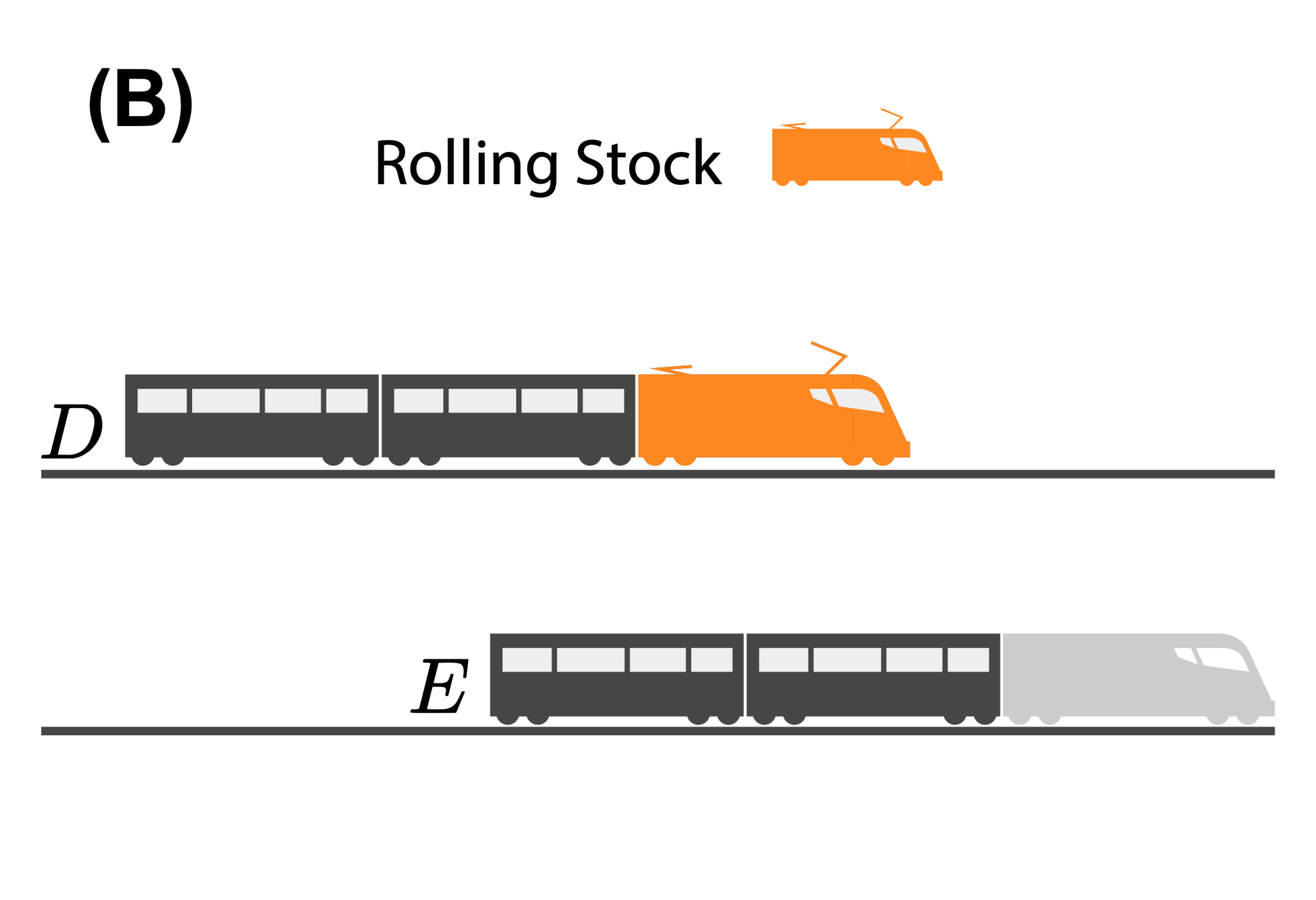}
\includegraphics[width=0.70\linewidth]{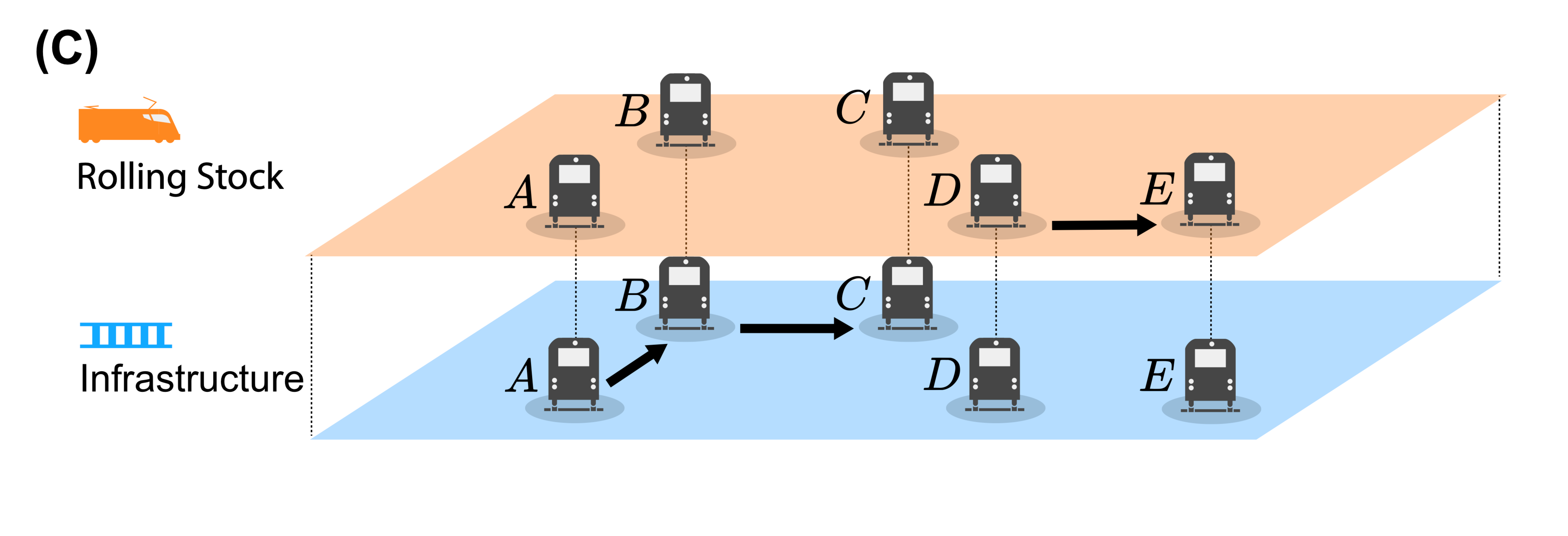}
\includegraphics[width=0.70\linewidth]{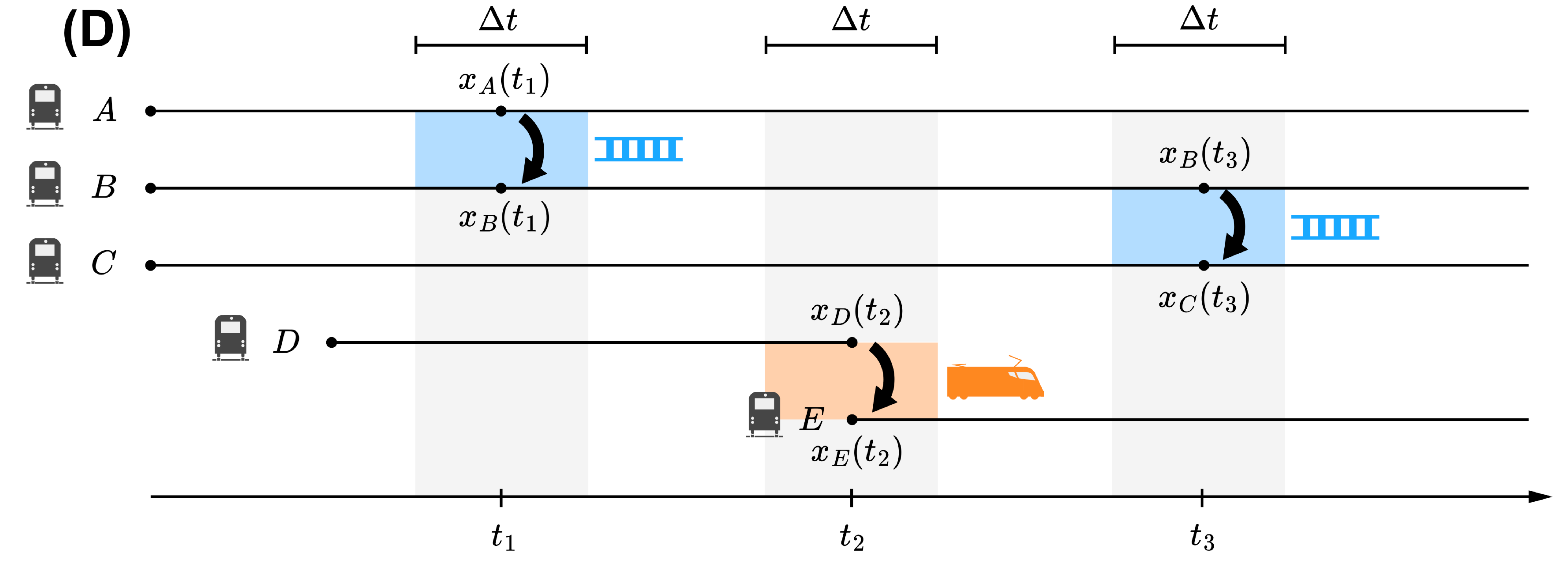}
\caption{\textbf{Multi-layer delay spreading.}
    Three different layers contribute to the spread of a delay.
    \emph{Infrastructure contact} (A) infrastructure (tracks or platforms) is occupied, e.g., by another train in transit.
    The upper train will then spread its delay onto the bottom train. In panel (C), we denote this by an arrow between $A$ and $B$ in the blue (infrastructure) plain.
    \emph{Rolling stock contact} (B) A train service has to wait for the arrival of the corresponding traction units being used by another service. The arriving train can spread the delay to the waiting train: in panel (C), train $D$ spreads the delay to $E$. \emph{Personnel contact}, not shown, where a train service has to wait for the onboard personnel to arrive.
    In panel (D), the bottom line represents the time axis.
    Train services are represented as the other lines. The beginning of a line represents the starting time of the service.
    A train service can transmit its delay through a \textit{proximity contact} (light blue) and a \textit{rolling stock contact} (orange), described in (C).
    In this picture, trains $A$ and $B$ compete for infrastructure resources (a free track section of line) at time $t_1$, so the status $x_{B}(t_1)$ (\emph{delay}) of train $B$ is updated through Eq.~\ref{eq:status}: all the contacts that $B$ had in the time interval, $\delta t$, are considered, and its delay is calculated accordingly.
    Trains $B$ and $C$ have a similar interaction at time $t_3$, propagating the delay generated from $A$.
    Train $D$ shares its rolling stock with train $E$ at time $t_2$, i.e., train $E$ needs to wait for train $D$'s arrival.}
\label{fig:multilayer}
\end{figure}

Considerable effort has been put into modeling the propagation of delays in transportation systems where analytical \cite{goverdeDelayPropagationAlgorithm2010}, agent-based \cite{gambardellaAgentbasedPlanningSimulation2002,monechiCongestionTransitionAir2015}, stochastic \cite{bukerStochasticModellingDelay2012,meester2007stochastic,pyrgiotis2013modelling,campanelli2014modeling}, networks~\cite{wei2015modeling,Sen_Dasgupta_Chatterjee_Sreeram_Mukherjee_Manna_2003, fleurquin2013systemic}, and purely data-driven models~\cite{oneto2017dynamic} were deployed. 
Railway networks have been modeled with agent-based simulations (ABS). They can be performed on different levels of detail. Macroscopic simulations, implementing only the queuing of trains for each available resource (blocks, platforms, rolling stocks, personnel), provide a rough - however intuitive and efficient approximation to delays in the network~\cite{zinser2018comparison,rossler2020simulation}. In contrast, microscopic simulations incorporate details of how a train moves and each of its interdependencies plays out, allowing for a more detailed simulation. Realism is gained at the expense of a more computational effort and the necessity of information~\cite{nash2004railroad,johansson2022microscopic}.

Simulating a national railway scenario is computationally challenging due to the amount of detailed information, such as multiple agents (trains), signals, splits, and primary delays. Computational efficiency is crucial as the number of details increases. For instance, a railway network simulation with 1,000 trains, 5,000 operational points, 100,000 signals, and 50,000 switches allows for billions of possible scenarios. These involve train conflicts, failures, queues, and delays. Traditional simulation approaches struggle with the associated combinatorial explosion caused by national railway networks, extensive tracks, numerous trains, millions of passengers, and complex scheduling algorithms. Incorporating every detail is limited by the associated  data requirements. Up to this day reasonable choices of which details to include and which ones not remain important. 

\begin{figure}[t!]
\centering
\includegraphics[width=0.80\textwidth]{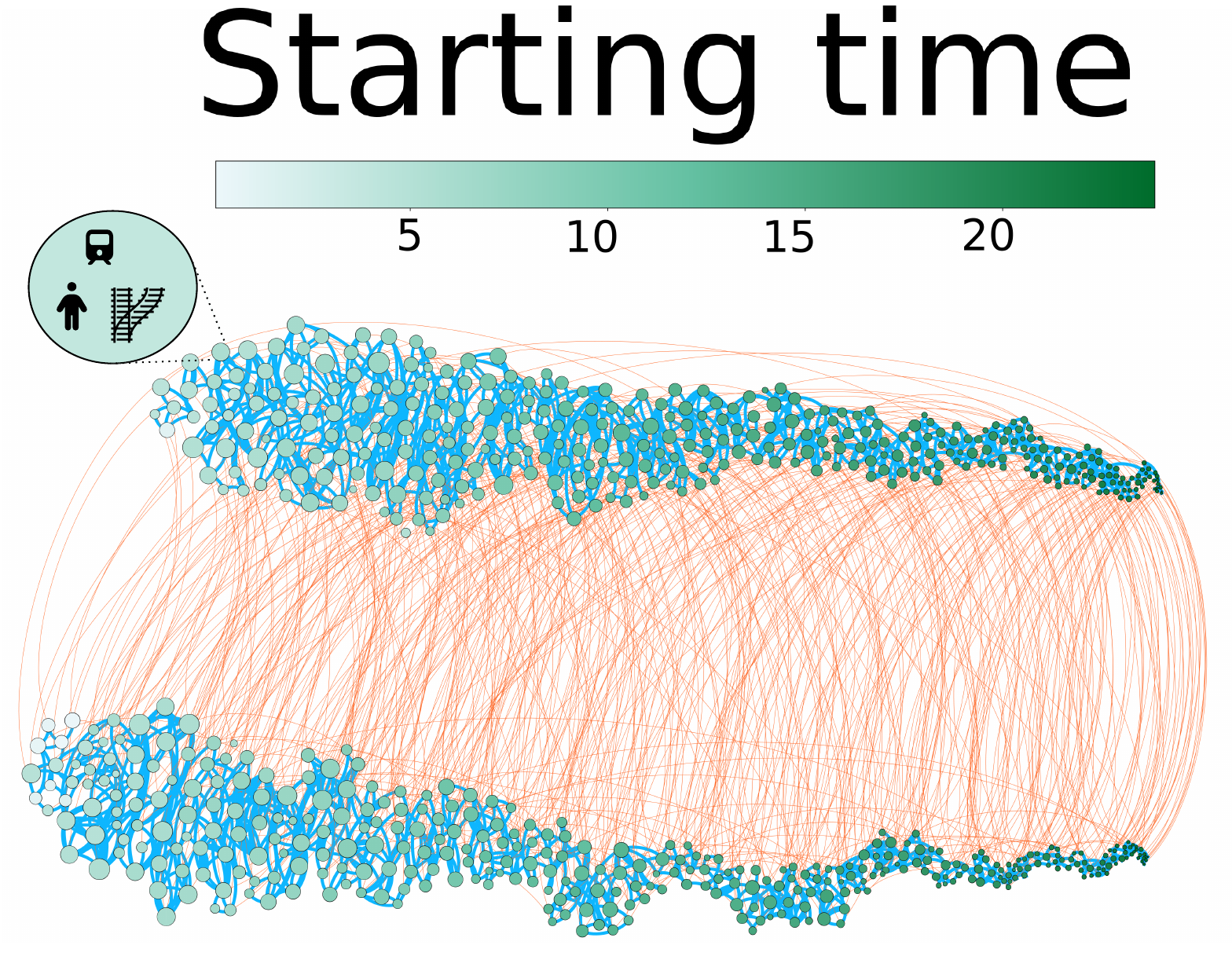}
\caption{
    \textbf{Effective impact network of train services} for a single train line between Wiener Neustadt and Vienna Central Station, for a typical working day.
    Nodes represent train services that contain three parts: personnel, rolling stock, and the availability of infrastructure (items in the circle).
    Orange links are interactions from shared rolling stock. Blue links show infrastructure contacts. The time of the day when the service starts flows from left to right, from ca.\ 4 a.m.\  to midnight. 
    The infrastructure contacts clearly define two communities that reflect the two travel directions. 
    The diameter of nodes is proportional to the impact of train services in terms of delay ``spreadability'' in the network.
    Note that trains operation in the first rush hours bears the highest impact.
\label{fig:startings}}
\end{figure}

In the railway network shown in Fig.~\ref{fig:multilayer} delay spreading can occur through one of two processes. First, two agents compete for the same infrastructure, i.e., tracks (blue layer). The time-ordering of events is important since, e.g., a train can spread delay to its following trains when it is delayed itself. Second, the delay may spread through interactions in the organization of the rolling stocks. If the traction unit of a running train is scheduled to be used by another train service on the same day, the potential delay of the first one may spread to the latter.
The two possibilities for interaction, represented by the blue and orange layers in Fig.~\ref{fig:multilayer}, create a \emph{multiplex} network that we call the {\em effective impact network}. The nodes are train services, links are the ways individual trains can negatively affect each other. 
Following Fig.~\ref{fig:multilayer}, if train $A$ has a delay, it will spread to $B$, and subsequently cascade to $C$;
Train E needs to wait for the traction units of train $D$, making the delay spread over the rolling stock layer of contacts.
The effective impact network for the path from Vienna Central Station to Wiener Neustadt in Austria is shown in Fig.~\ref{fig:startings}. 

\begin{figure*}[ht]
\centering
\includegraphics[width=0.95\textwidth]{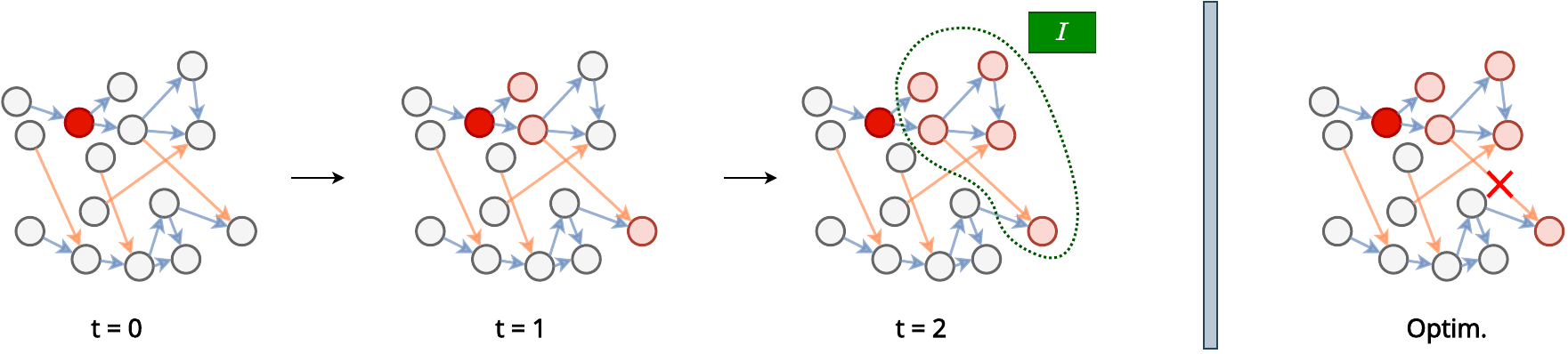}
\caption{ 
\textbf{Network diffusion procedure (NDP) and Impact Reduction.}
    We show the steps for reducing delay spreading in the railway network. 
    For every train, we start a diffusion at $t=0$, updating its status (delay) to $1$. 
    The train then transmits its delay to all its trains downward the effective impact network ($t=1$). 
    At the end of the procedure, once all lower trains are impacted (here at $t=2$), the sum of the delay of each train in the network represents the \emph{impact}, $I$, as defined in Eq.~\ref{eq:impact}. 
    Through a simple network re-organization (here removal of an orange link), an optimization technique (\emph{Optim.}) leads to less expected delay in the system.
\label{fig:impact}}
\end{figure*}

{\em Network diffusion procedure.}
The main idea of the present work is to use the effective impact network to compute a systemic risk indicator, the {\em train impact}, $I$, for every train in the system. $I$ is defined (see {\bl Methods} in Eq.~(\ref{eq:impact}))  as the sum of the delays transmitted to other trains at the end of a network diffusion procedure (NDP) that we sketch in Fig.~\ref{fig:impact}.
For every train, we start a diffusion process at some internal time $t=0$.
Initially, all trains start with no delay. We then pick one train and assign a delay of $1$ to it. The train then transmits its delay to all trains that lie one step downward the effective impact network ($t=1$). This procedure repeats until all downward trains are impacted (here at $t=2$). The sum of the delay of all trains in the network represents the \emph{impact}, $I$, as defined in Eq.~\ref{eq:impact}. 
In other words, based on this network we design a network centrality measure corresponding to a specific diffusion process that is a realization of the possible cascading processes.  
We use the trains' systemic impact to rank the systemic relevance of train services. 
By modifying the network, e.g., by introducing new train services we can change the impact of trains and thus the ``spreadability'' of delay in the system. 
Introducing new train services typically means reducing dependencies of trains and is equivalent to pruning links of the initial effective impact network (see  Fig.~\ref{fig:impact} D).

{\em ABS.}
To validate the effectiveness of the results obtained with the network diffusion procedure (NDP) we develop a corresponding ABS that captures the delay distribution of the system; see {\bl SI text S1}.
We can now show that delays in the system can be significantly reduced by only adding a few specific new train services (that amount to deleting links in the effective impact network). 
The NDP is particularly convenient to test railway management policies without detailed data or computationally exhaustive simulations.

In the following, we primarily focus on the Austrian railway section from Vienna Central Station to Wiener Neustadt, a small town at approximately \ 40 km south of Vienna. This section is part of the S\"udbahn line, one of the most traveled lines in Austria. In the SI, we provide results for the whole of Austria and another selected line (see {\bl SI text S2,S3}, respectively Fig.~{\bl S2} and Fig.~{\bl S3}).
In {\bl SI text S4}, we provide a glossary of railway terms used.

\section*{Results}
\label{section:results}

In Fig.~\ref{fig:startings} note the multiplex nature of the contacts in the railway network. The size of the nodes is proportional to the train's impact in initiating delay cascades. The color encodes the trains' starting time. We observe that the most impactful trains are those running during the first rush hours (larger circles).
The two communities of nodes (the upper and lower strains) come from the topology of the infrastructure contacts (blue links) and interact through the rolling stock contacts (orange links).
The different types of contacts (see {\bl SI text S5} Fig.~{\bl S4}) spread delays at different rates.
Even though the delay transmission in the infrastructure layer appears to be more frequent, nodes are affected (get delayed) with less intensity than through the rolling stock layer. 

Note that in our analysis, we do not consider the circulation (reassignments) of railway personnel (drivers, conductors, inspectors) due to a lack of corresponding data and that the diffusion method reflects the different spreading strengths in the two layers with two different parameters (see {\bl Methods}).

\begin{figure*}[ht!]
\centering
    \includegraphics[width=0.55\linewidth]{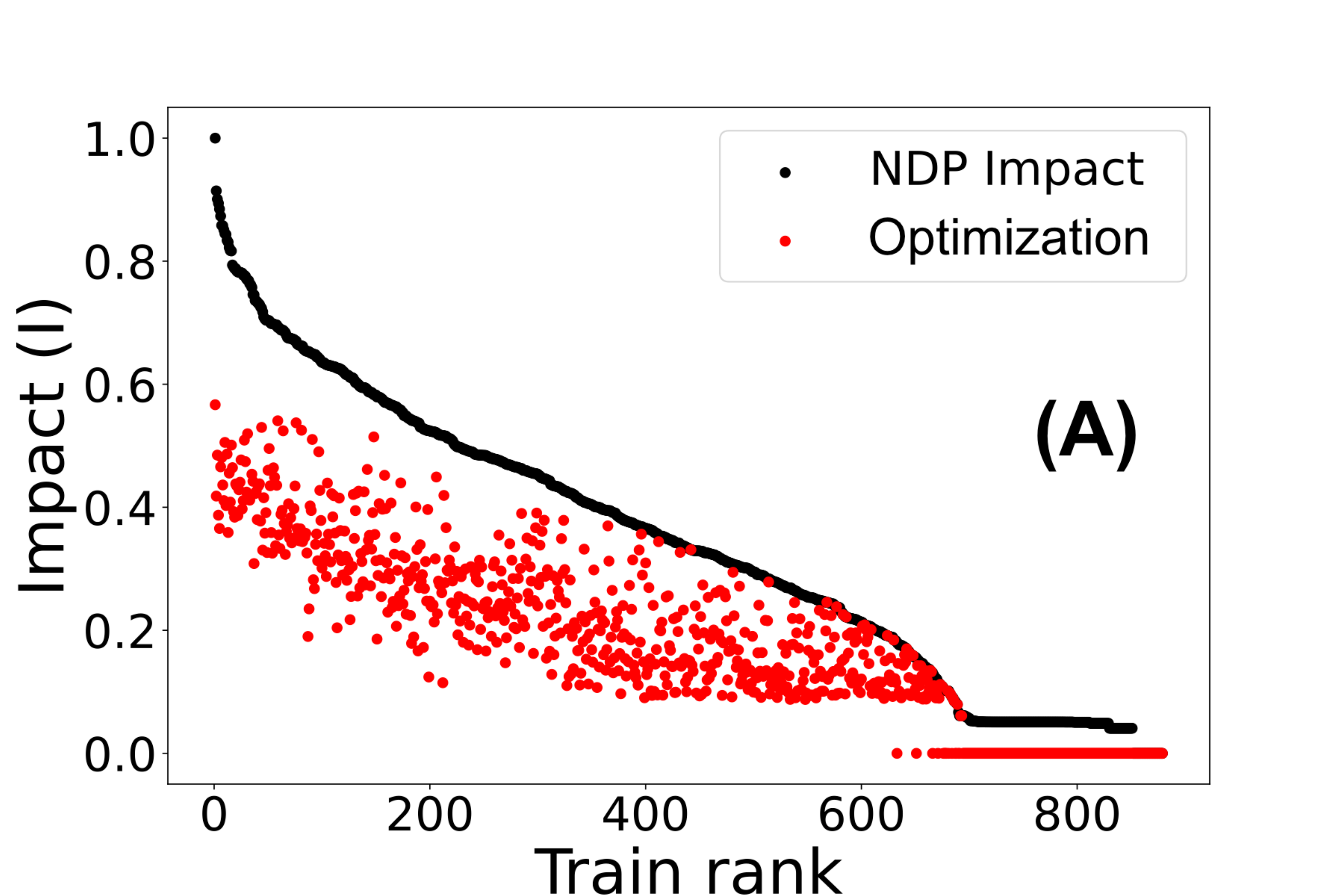}
    \includegraphics[width=0.41\linewidth]{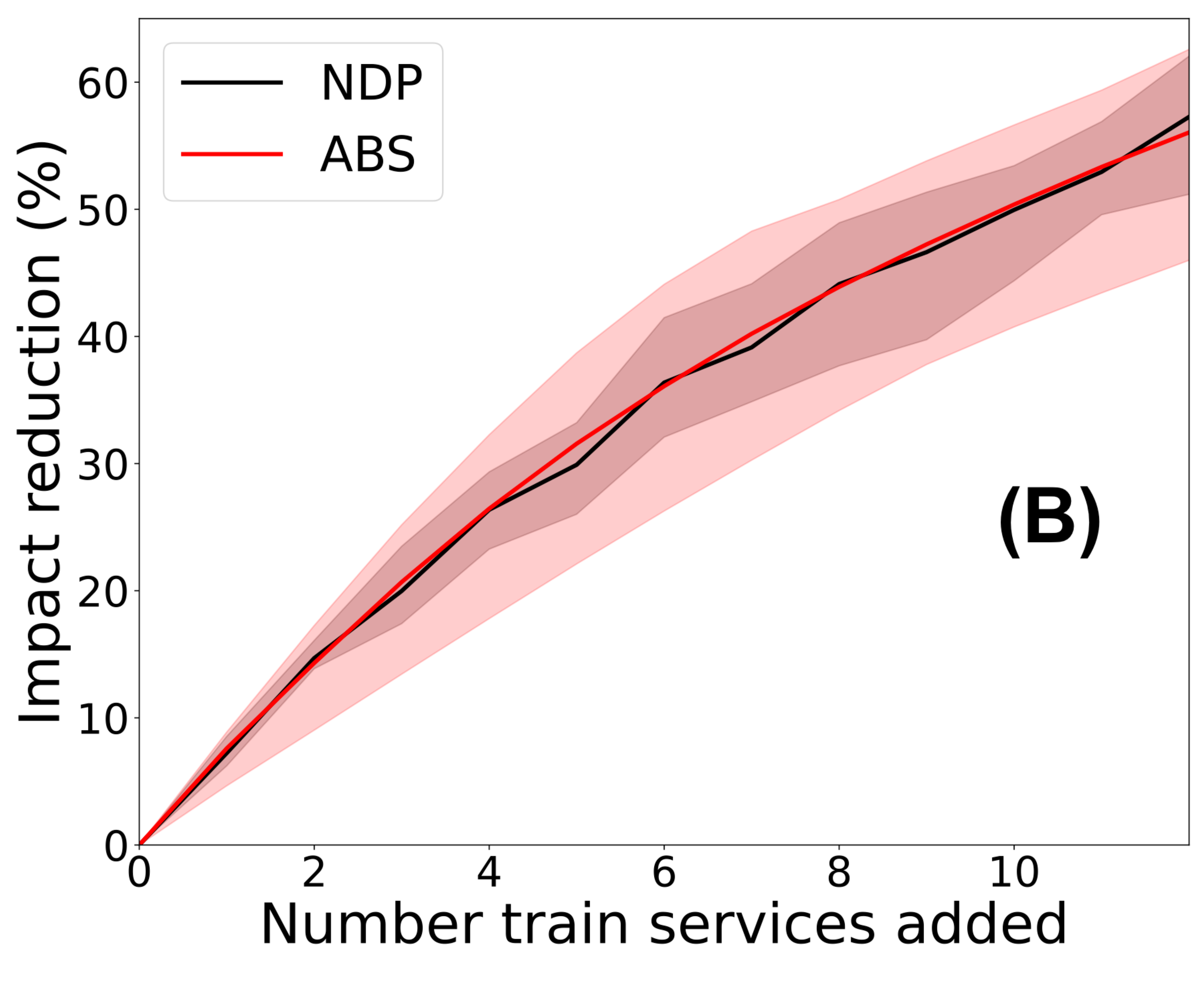}
    \includegraphics[width=0.50\linewidth]{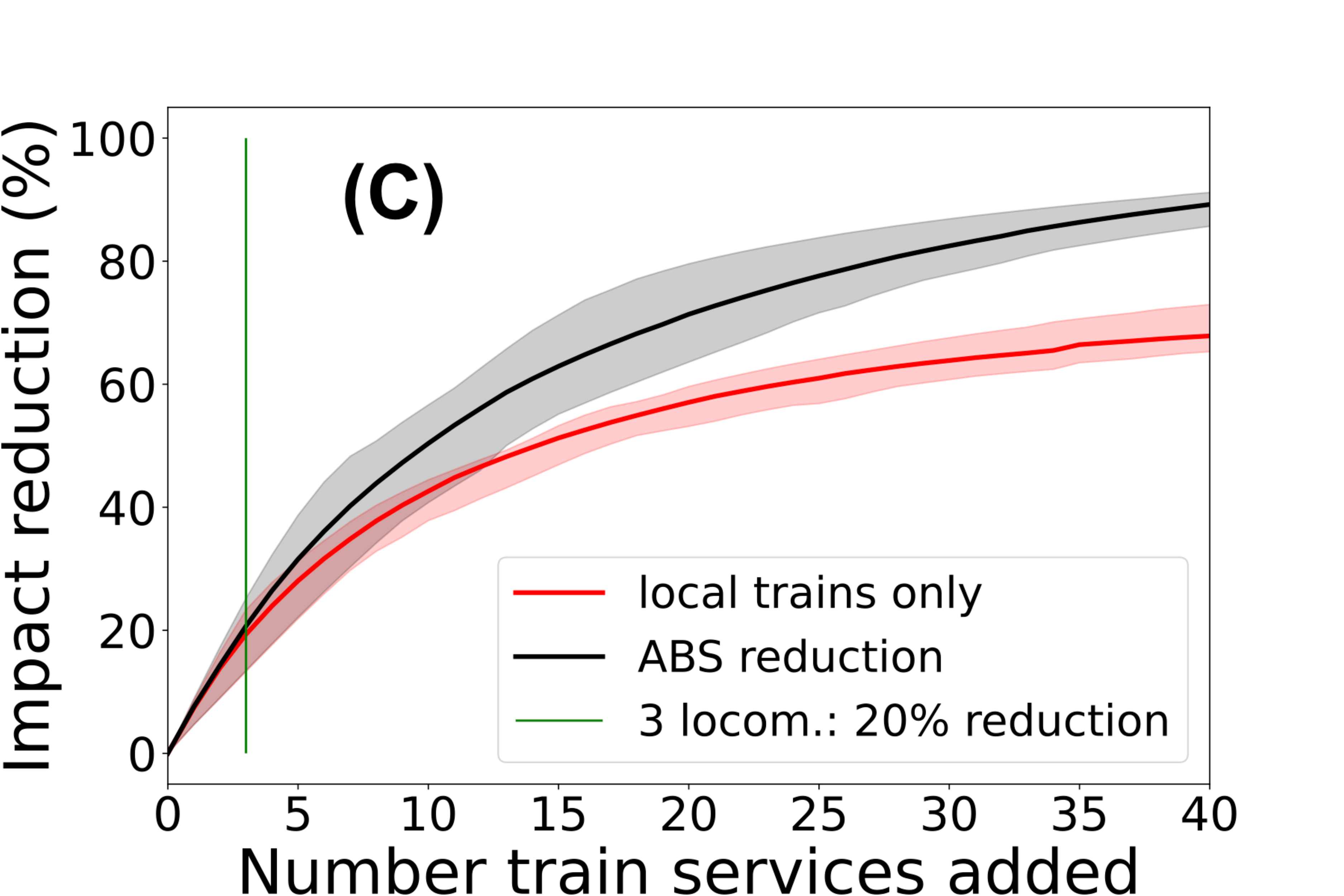}
\caption{\textbf{(A) Ranked train impact}, $I$. 
        The black line represents the results of the NDP with both layers, infrastructure, and rolling stock. 
        Red points consider the infrastructure network only and
        are ordered according to the original rank (black line).
        \textbf{(B) Delay reduction} by adding new train services.
    Average, first, and third quantiles of the relative impact reduction as a result of adding new train services.
    For the ABS (red) the impact is in terms of total delay transmitted, the NDP is in black.
    NDP uses a recursive algorithm in the network of contacts to assess which rolling stock would have the highest impact on sharing delay.
    Based on that, we progressively add more rolling stock to reduce their impact and break some of the black links in Fig.~\ref{fig:startings}. 
    With three more rolling stocks we reduce the impact by about \ 20\% (green vertical line).
    For the ABS we find that introducing three more units decreases the total delay of the day by \ ~ 20\% (as also found with the diffusion method) while introducing 20 units more reduces delays by about \ 65\%.
    Lines converge at\ 100 \%.
    \textbf{(C) Delay reduction by specifically adding local train services.}
    The ABS (black) covers all the possible train services, which coincides with the green curve in panel (B).
    The red curve is obtained analogously to the black one but by restricting the procedure to local train services only, i.e., the cheapest and easiest to add.
    While the black curve eventually reaches 100\%, the red one saturates at 70\%.
    Note that with three more rolling stocks we reduce the impact by approximately \ 20\% (green vertical line).  }
\label{fig:rankplot}
\end{figure*}

In Fig.~\ref{fig:rankplot} we show the ranked distribution of the train impacts  with (black line) and without the rolling stock contacts  (red dots) for the NDP. The latter case represents the idealized situation where rolling stock would not have any effect on delays. 
For comparison, we compute an accurate fully detailed ABS, described in the {\bl Methods} section and sketched in Fig.~{\bl S1} (A) in {\bl SI text S1}, as a benchmark. Fig.~{\bl S2} (B) in {\bl SI text S1}, shows how well the simulation manages to reproduce trains' delay at destination during a whole month.

A corresponding rank distribution of the impact has been produced with the ABS and is shown in {\bl SI text S6}, Fig.~{\bl S8}.
In both cases, NDM and ABS, we observe a substantial reduction of the impact of the most dangerous trains by about a factor of\ 2.  

We continue with case-based scenarios, deliberately introducing primary delays to a small fraction of the most impactful trains to assess their dangerousity and coping for the cascades generated. 
We use NDP and ABS to asses the impact reduction in a scenario in which the top 2\% most impactful trains are the ones generating cascades (\emph{influencer trains}).
In the ABS, we inject a one-hour primary delay on the initial trains; in the NDP we start the diffusion from all of them, assigning a value of $1$ to their node status.
In the following, we refer to the impact reduction generated by multiple initial trains, as \emph{multi-body impact}.

Figure~\ref{fig:rankplot} (B) shows the delay reduction percentage by gradually adding train services according to their contribution to their multi-body impact. Note that adding train services is equivalent to pruning the rolling stock contacts.
For the ABS we interpret the overall transmitted delays as the impact.
Adding three train services already reduces the total delay by $\simeq$ 20\%. 
We obtain similar results in scenarios where the amount of train influencers ranges between 1 and 5\%. More than 5\% would be far from  realistic.
The red curve represents this optimization for the ABS.
The black line represent the results of the diffusion over our network.
The two methods show agreement for the first dozen of added trains. 
The optimization procedure has been performed over all the possible train services. However, long-distance trains are more difficult and more expensive to be substituted. In Fig.~\ref{fig:rankplot} (C) we show the optimization using the ABS in two cases:
(i) Optimizing over all possible train services (black line) and (ii) optimizing over only the train services that are the cheapest for the railway companies to add (red line); these are the local trains~\cite{nakayama2018optically} with electric traction units.
In this way, we show that we can reduce the overall delay by about the same percentage (20\%) by adding three of the cheapest possible train services that reduce the effects of local disturbances.
The results are shown for a quarter (from January 1$^{st}$to March 31$^{st}$ 2019) in the Austrian S\"udbahn railway line from Vienna Central Station to Wiener Neustadt.

\begin{figure}[t]
\centering
\includegraphics[width=0.70\textwidth]{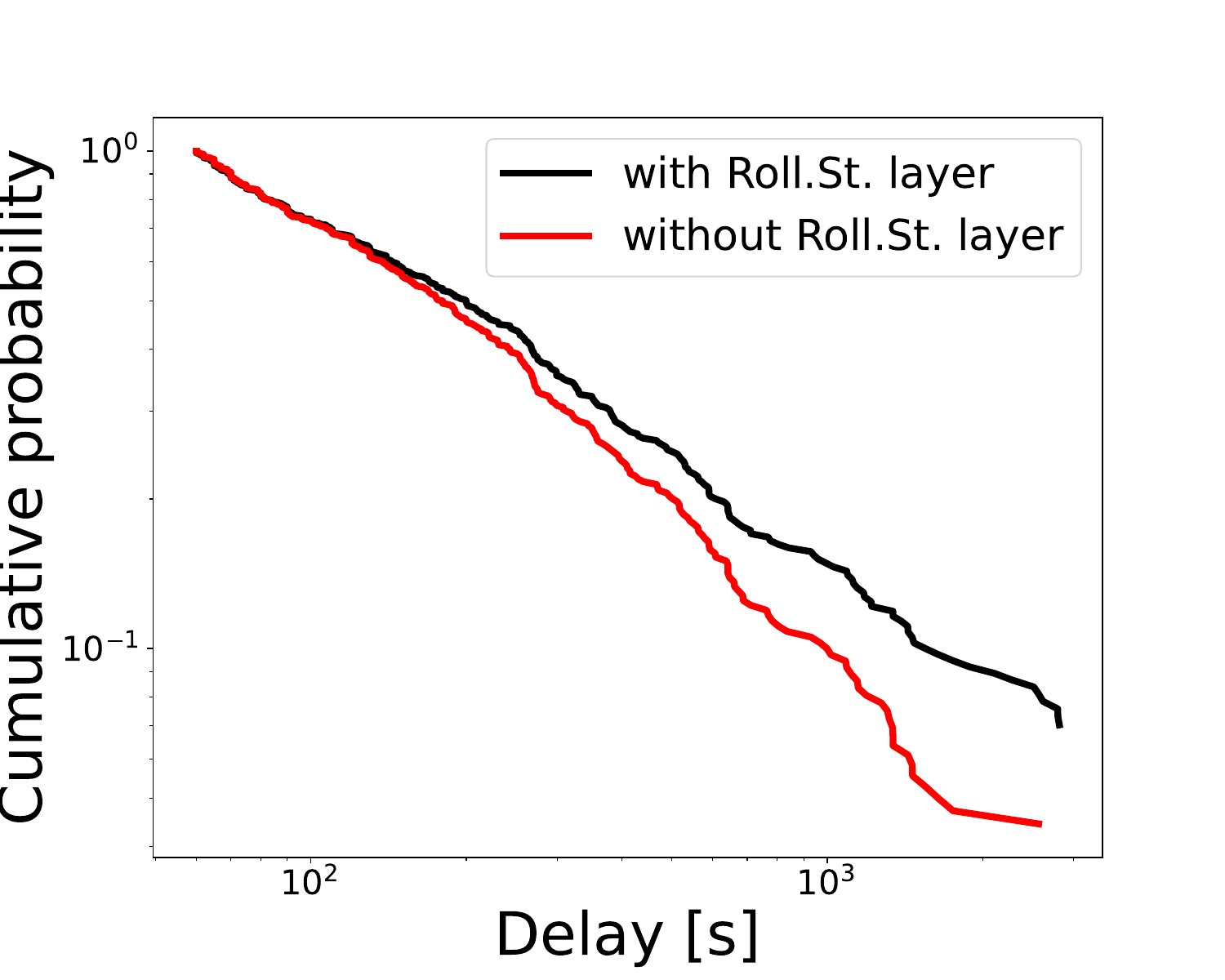}
\caption{ 
\textbf{Change in daily delay distributions.}
    The black line represents the delay distribution of the 2\% scenario for the entire month of May 2018. 
    After removing all  rolling stock dependencies we observe a substantial reduction in the tail of the distribution (red).
\label{fig:tail}}
\end{figure}

In Fig.~\ref{fig:tail} we show the change in daily delay distributions (at the final destination) when removing all rolling stock dependencies. We use the scenario where the 2\% most impactful trains get an hour of primary delay. The corresponding delay distribution is the black line.
When all trains are independent of the rolling stock layer we get the red line. A substantial reduction in severe delays of almost a factor of six is seen. Note that 60\% of this reduction is reached for the first \ 12 steps of the procedure.

\section*{Discussion}

We use a novel approach capable of significantly reducing the size of delay cascades in the railway system.
We analyzed the topology of the network of train contacts (Fig.~\ref{fig:startings}) by which trains may induce delays on other trains, calculating the impact for each train.
We identified rolling stock sharing as the main factor responsible for delay cascades. 
This happens despite the number of links in the rolling stock layer being way smaller than that of the infrastructure contacts (see {\bl SI text S5} Fig.~{\bl S4}).
The impact of each train and the densities of the contacts for each layer suggests that to mitigate delay cascades, one should focus on strategies affecting the supply constraint layers (rolling stock or personnel). 

If we only consider  infrastructure contacts, railways are resilient as soon as the system is far from being overloaded. 
Effectively managing the organizational layer can help ensure the overall resilience and stability of the railway system, even in the presence of unexpected disruptions.
This insight can provide guidance in the details that a railway agent-based simulation needs to incorporate in order to correctly simulate delay cascading.

Our Network Diffusion Process (NDP) is effective in ranking trains by their propensity to spread delay, an essential insight to start coping with the highest delays.
In practice, if we have two big train influencers, maybe not both of them are dangerous in reality.
For example, a  topologically very central node never spreads anything because the corresponding train service is always on time. Here we must specify the difference in characterizing the topology of a network or a phenomenon happening \emph{on} the network. A realization of a random event such as delay arising in the railway network and its consequent spreading is a phenomenon happening \emph{on} the structure of the contacts.
For this reason, we wanted to rely on a more detailed simulation that could validate our results.
As a benchmark, we devised an agent-based simulation (ABS) to reproduce the Austrian railway's daily dynamics.

In a scenario with no delays, i.e., with no conflicts, both the ABS  and the NDP measure scale linearly in the number of trains and operational points.
When delays are present, the ABS resolves all the conflicts generated, thus adding extra computational time needed to allocate all the extra events. 
The temporal network diffusion will not explicitly take into account all these conflicts. 
In this sense, the diffusion technique can be seen as the conflict-static representation of the whole ABS, resulting in a more pragmatic but still effective characterization of the system.
Simulating the system when large delays happen, namely when conflicts happen, is a difficult task. 
A lot of microscopic details are needed to fully simulate the system and obtain useful insights.
The network measure has the advantage of not depending on these details, thus it is cheaper in terms of data and needed data accuracy.

Moreover, sensors used to record train trajectories may fail, generating inconsistencies in the acquired data. 
Fixing data collection bugs, in fact, is an actual field of research \cite{barbour2020enhanced}.
Leveraging NDP, we have been able to improve the system's overall performance with a simple strategy consisting of adding a few resources. As an example, we show that this is possible by introducing additional train services, avoiding the interdependencies in crucial events during the day. 
The diffusion method described in \cite{dekker2022modelling} focuses on a network of stations, while our nodes are trains, i.e., the real delay diffusing agents.
Our diffusion technique is rather simple, but, as shown with the ABS, captures the essential part of delay cascading.
Our impact is only a proxy for the delay, being a topological measure.
However, through the ABS we showed that results coincide with a certain precision.
For a more detailed description of the precision, we refer to Fig.~{\bl S9} in {\bl SI text S7}.
Moreover, our approach is lacking the personnel layer, which could in principle change the shape of trains' impact distribution.
If the diffusion technique, based on the impact, is significant for other national railways is out of scope here.
This approach is quite different from diffusive studies on railway networks, being that focuses on train services (and not stations or operational points) as nodes of the network.

We defined an optimization procedure able to reduce the impact on each railway line.
Contacts between lines are significantly lower than the ones inside a line (see Fig.~{\bl S5} in {\bl SI text S5}), validating the compartmentalization of the national railway.
In this way, we are neglecting those contacts, which could in principle be re-introduced as a second-order effect.
By ranking trains' impact we unveiled the fact that some trains are more dangerous for the whole system's performance, and this is due to the rolling stock layer, accounting for half of the total delay.
If we only consider the infrastructure layer, no trains are notably big influencers.
This approach ranks trains based on their single-body impact, without taking into account the joint disturbance generated by multiple trains.
We performed the same ranking based on the total delay induced in the ABS, as shown in Fig.~{\bl S8} in {\bl SI text S6}.
{\bl Figure~\ref{fig:rankplot} (B)} demonstrates how we resolve a concrete problem, i.e., reducing the size of the delay cascade in a scenario in which some trains gain an important delay.
By adding just three new train services, i.e., by removing three train dependencies, we achieve a 20\% reduction in the total delay at the end of the day. 
The two approaches, NDP and ABS agree on the first and most important steps of the optimization curve.
Adding more trains the two methods differ and the NDP overestimates the impact reduction. A possible explanation is that the ABS contains far more details, including the buffers at each station that generally reduce the impact. However, particular combinations of the two parameters in the diffusion simulation can accurately reproduce the impact reduction estimation of the ABS; see {\bl SI text S7} Fig.~{\bl S9}.
Of course, we have a reduction when a critical situation happens: when each train is on time we would observe no reduction.
We performed the same optimization over another line, line 10102 from Vienna to Salzburg, Fig.~{\bl S2} in {\bl SI text S2} and over the whole Austrian railway network Fig.~{\bl S3} in {\bl SI text S3}.
For this last optimization, we also included the contacts between lines, showing that the second-order contacts do not inhibit our NDP optimization.
We observe that on the line with less traffic, the optimization is more effective, reaching almost 60\% impact reduction by adding only three train services.
For the whole Austrian network, the total reduction amounts to 40\%, adding the same percentage of trains with respect to the total, which in this case amounts to 37 trains.

We also restricted the optimization over the local trains, which are the easiest and cheapest to reassign.
In this case, we can still reduce the delay by 20\% just by adding three train services {\bl (Fig.~\ref{fig:rankplot} (C)}).

Moreover, this procedure reduces the probability of the largest delays by a factor of six, {\bl as shown in Fig.~\ref{fig:tail}}.
This is the distribution we would obtain pruning all the rolling stock layer, so adding hundreds of trains.
Nonetheless, 60\% of the discrepancy between the two curves is made by adding only 12 trains.

With respect to Ref.~\cite{ball2016two}, we simulate a real situation and do not investigate the load of trains at the stations and the collapse of the system with a mean-field approach.
We, instead, quantify the spreading ability through the rolling stock layer.

The central piece of this work is to introduce a metric of systemic risk and use it, in a policy-making scenario, to reduce trains' impact on the whole system's delay.
To achieve this goal, we used a network made of train services.
Reducing the impact is not as reducing the total delay, as the impact in our NDP is a topological measure over the contact network.
To validate the results, we used a more precise ABS, showing that our procedure reduces the impact over the network as would reduce the total delay in an ABS.

We could study the contacts between the lines and understand which importance they have onto the single-line dynamics.
One important tool would be a measure of early warning signals of disruptive transition: a measure dependent on all the layers able to anticipate and control train traffic flows.
A helpful insight would be to weigh these procedures with energy consumption, carbon dioxide reduction, and the number of passengers, but these pieces of information are currently missing.

An important feature to add in future studies would be personnel circulation. 
The complete picture containing all layers could add more depth to the research and topic.
Using this insight, it would be possible to generate a new sort of policy, rewiring the multi-layer connections between trains, instead of pruning them.
This would result in a long-term rescheduling of trains, instead of adding new ones.
This could greatly reduce cascades and total delays on a daily just by re-organizing the system in order to be resilient to delay cascades.


\section*{Materials and Methods}

\subsection*{Data}
\label{subsec:data}

The Austrian Railways, \"OBB, provided the data used in this work.
Due to the sensitivity of these data \"OBB agreed to share only one week of train operations (real, scheduled, and rolling stock sharing) in a specific anonymized form. With these data, containing the scheduled, real-time, and rolling stock sharing, it is possible to reproduce the network results of the article.
We collect the primary delays to inject into the ABS from an annotated delay data frame provided by \"OBB.

\subsection*{Methods}

The subject of analysis is \textit{train services}. Train service is the collection of services and resources needed for a train to run, e.g., rolling stock, infrastructure, and personnel. In the public daily schedule, train services are indicated by a unique identifier, e.g., ``RJ 123'' (train class: Rail Jet; train number: 123).
We define two types of interactions or contacts between trains:
\textit{rolling stock contacts} and \textit{infrastructure contacts}. 
Rolling stock contacts are responsible for the transmission of delay caused by shared rolling stocks (wagons, traction units), e.g., that are not at the right place at the right time when needed. 
Infrastructure contacts may cause transmission of delays by shared facilities (platforms at stations, sections of lines, and tracks).
More precisely, a train that will pass by a block has a direct connection to the next one that will pass by the same block.

{\bf Effective impact network.} We consider the railway system as a multilayer network with nodes representing train services and links in each layer representing one of the two types of contact. Each line of the Austrian national railway service is split into two main components, i.e., the two opposite directions of the line (Fig~\ref{fig:startings}).
The rolling stock layer bridges these two components that would remain disconnected otherwise. 

With these premises, we compute diffusion processes on the \textit{multiplex network} for all the train services.
Dynamical network processes can be of different nature~\cite{Borgatti_2005}. The closest to the phenomenon of interest is diffusion via the replication mechanism. It has been used e.g. to simulate \textit{influence}: through interaction, individuals cause changes in each other’s beliefs or attributes~\cite{sajjadi2022structural,barrat2008dynamical}.
Performing this action, the influencer does not lose his/her attitude and can influence multiple people simultaneously.
Similarly, a train influences another train without losing its delay and keeps its capability to transmit delay. Multiplex networks are employed to model scenarios where the same class of nodes interact in various ways, such as in transport systems. 
By assigning distinct colors to links of various types, multiplex networks can be visually represented as a colored network in one layer. For example, in Fig~\ref{fig:startings}, we can see how the nodes (train services) densely interact with infrastructure contacts (blue edges) inside each of the two major components; instead, the two strands are connected by rolling stock contacts (orange edges).
In the multiplex formalism, the same node $(i)$ has a \textit{replica node} $(i,\alpha)$ for each layer $\alpha$.
In this notation, the diffusion equation in a multiplex network is
\begin{align}\label{eq:diff}
\frac{d x_i^{[\alpha]}}{d t} &= D^{[\alpha]} \sum_{j=1}^N w_{i j}^{[\alpha]}\left(x_j^{[\alpha]}-x_i^{[\alpha]}\right)+ \nonumber \\
& + \sum_{\beta=1}^M D^{[\alpha, \beta]}\left(x_i^{[\beta]}-x_i^{[\alpha]}\right) \quad ,
\end{align}
where $x_i$ represents the status or attribute of the node $i$. In our case, it is a proxy for the delay. The first and second terms on the right-hand side of the equation correspond to intra-layer and inter-layer diffusion, respectively.

In this network framework, we can compute the so-called \textit{communicability}, i.e., the number of reachable nodes from the starting node, through a diffusion process with infinitely many steps.
Each train is connected to the network section corresponding to train services running at a later time. From a communicability point of view, each train could in principle reach each other train starting in a following moment in time. For this reason, we analyzed the speed of the diffusion process.
Since we want to address the risky state of train services, the second part of Eq.~(\ref{eq:diff}) coincides with $x_i^{[\alpha]}$ and 
$D^{[\alpha]} w_{i j}^{[\alpha]} = w^{[\alpha]}$, since the coefficients are only layer dependent. 
In line with a previous study~\cite{goverdeRailwayTimetableStability2007} we use \textit{max-plus} operations to update each node state.
The final update equation is:
\begin{align}\label{eq:status}
x_i^{[\alpha]}(t+\Delta t) &=  \max_{\{j=1...N\}}\left(w^{[\alpha]} x_j^{[\alpha]}(t),x_i^{[\alpha]}(t)\right) \quad ,
\end{align}
where $j$ runs over all the contacts agent $i$ had in the time interval $[t, t + \Delta t]$. In this sense, our diffusion measure of delay spreading depends only on the network topology. 

We carried out a type of diffusion in which a small amount of delay (e.g., $w^{[\mathrm{roll.st.}]} \approx 0.2$) gets transmitted through infrastructure contacts while rolling stock contacts transmit a more significant amount ($w^{[\mathrm{infrastructure}]}\approx 0.7$). The material constraints play an essential role in the diffusion of delay(see {\bl SI text S5} Figs.~{\bl S6, S7}). We simulated the impact of each train service by initializing the corresponding node with a constant status, i.e., the status of node $i$ in every layer $\alpha$ at the initial time $t_0$ is set to $x_i^{[\alpha]} (t_0)=1$,  and letting it diffuse on the whole network.
All the other trains are initialized to $x_j^{[\alpha]} (t_0)=0$.
We calculate the train service's impact by summing all nodes' states after the diffusion process has converged. We calculate the \textit{impact} $I$ of a train as
\begin{equation}
\label{eq:impact}
I = \sum_j x_j^{[\alpha]} (t_\mathrm{f}) \quad ,
\end{equation}
where $t_\mathrm{f}$ is the time of the last event of the contact sequence. In the case of ABS, we identify the overall transmitted delays as the train's impact. As shown in Fig.~\ref{fig:startings}, these diffusion processes show a backbone of more impactful trains whose starting time is in the first rush hours. If we only consider infrastructure contacts, we will not observe this effect.

If only the infrastructure layer is considered, most of the first rush hour delays do not propagate to the second rush hour. By adding the less frequent rolling stock contacts, the diffusion on the network will happen faster, quickly covering the whole network. As shown in the Dutch Railways in previous studies~\cite{dekker2021cascading}, these findings highlight how the constraints in the available rolling stock materials are essential for the delay spreading phenomenon and must be considered.

Once we quantify the risk of delay spreading associated with all the train services, we perform a recursive algorithm where we optimize the addition of new train services to break the cascade of delays.

Since the diffusion of delays on the infrastructure layer is slower than on the rolling stock layer we shall determine which rolling stock link to break when optimally adding new services to minimize delay cascade sizes.

We generate a scenario in which the high-impact trains gain some important delay. We simulate a diffusion starting from the most dangerous trains. We vary the number of starting trains from 1\% to 5\% of the entire number of trains (around \ 1200 trains), observing comparable results.

The optimization works as follows: we initialize the status of the most dangerous trains, for example, around $1-5\%$ of the total number of trains. Each value generates a scenario that follows similar dynamics; in the following, we will refer to the choice of $2\%$. Then, we quantify the impact of the diffusion simulation on the whole system. Subsequently, we optimally add a train service to reduce this impact maximally. 
In practice, this means pruning the rolling stock contacts, removing first the most dangerous in terms of impact. Few rolling stock contacts play a central role in the diffusion.


\section*{Acknowledgments}
We acknowledge the Austrian Federal Railway (\"OBB) for funding this work under framework agreement \emph{Train Operating Forecasting}, agreement number 341/4600029654. We are grateful to the CSH Visualization Team, Liuhuaying Yang and Tobias Batik, for their help in developing the explanatory images used in the article. We acknowledge insightful discussions about railway data and simulations with Michael Holakovsky and Matthias Wastian.


\newpage

\section*{Text S1: The Agent-Based Simulation (ABS)}

We devised a macroscopic agent-based simulation of the Austrian railway to have a more intuitive and quantitative measure of trains' impact and test if our topological diffusion measure can give hints on an actual situation.

\begin{figure*}[ht!]
\centering
\includegraphics[width=0.49\linewidth]{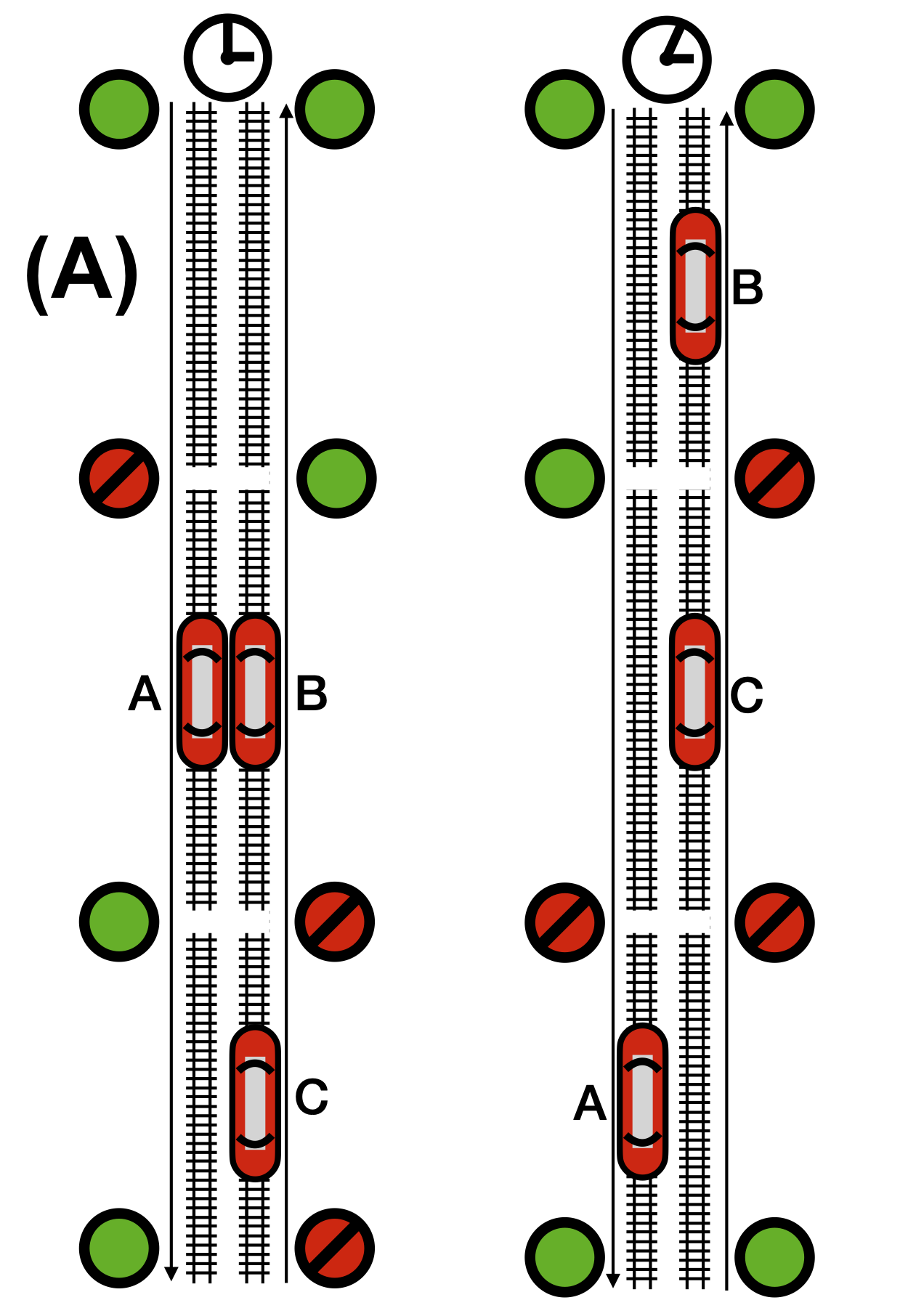}
\raisebox{0.3\height}{\includegraphics[width=0.49\linewidth]{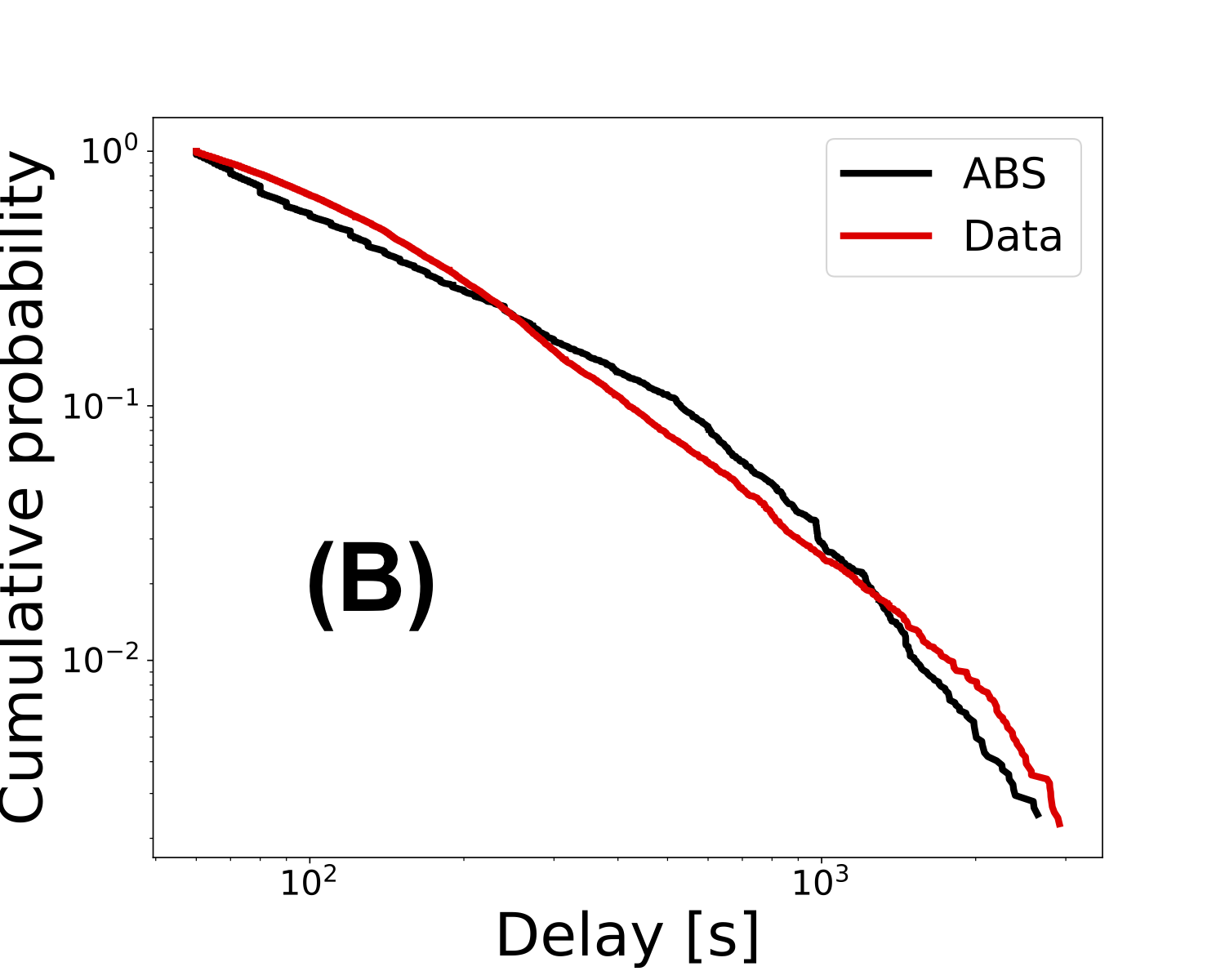}}
\caption{\textbf{Simulation vs.\ real.}
(A) Visual representation of the agent-based simulation at two different timesteps.
We depict eight signals (full green circles are ``go'' signals, red barred circles are stop signals), six blocks between signals (we separate them from each other by a small gap for visualization purposes), and three trains (A, B, C).
Arrows represent the traveling direction.
If a block is occupied by a train, the signal at its border is set to red, and trains must wait for its clearance, e.g., train C on the left side of the cartoon.
Our macro-simulation treats the whole system as a set of queues, one for each block.
(B)
We compare the cumulative distribution of the delays at the final destination resulting from our simulation and determined from historical data.
The delays at the final destination include the effects of primary and secondary delays.
Data refer to May 2018. 
\label{fig:exo_distribution}}
\end{figure*}

We built a \textit{macroscopic agent based simulation}, available as a \href{https://github.com/vitelot/training}{GitHub repository}.
Our simulation handles an event-based queueing system where each train only travels part of its trajectory when the needed services (infrastructure and rolling stock) are available. 
We sketched the basic mechanism in Fig.~\ref{fig:exo_distribution} (A). 
We differentiate between \emph{exogenous} delays (also known as primary delays) from \emph{endogenous} (secondary) delays. Exogenous delays are unpredictable and are caused by factors external to the railway system. Typical examples are bad weather conditions, strikes, accidents, and crowd congestion. Endogenous delays occur due to trains interacting in the system, e.g., a train travels slower than expected on a track and stops the regular flow of other trains. We can determine the frequency distribution of exogenous delays using delay data annotated by train conductors and drivers. This distribution is fat-tailed with a natural cut-off of 24 hours.

The simulation works as follows. First, we select the day to simulate and fetch the corresponding scheduled timetable. Then, we sample exogenous delays on trains at given locations based on their historical occurrence and superpose these delays to the timetable.
The simulation accordingly reproduces endogenous delays.
In Fig.~\ref{fig:exo_distribution} (B), we showed the distributions of trains' delays generated by the simulation against the real ones at the end of the day. We can evaluate the relative risk of causing endogenous delays associated with different trains by simulating a typical day. 
Specifically, we simulate a typical day with no massive delays and intentionally introduce a considerable delay for each train, one by one. This procedure requires running as many simulations as the number of trains in the system considered, e.g., a single line. We rank trains according to their impact in terms of delay caused to other trains. We use this information to simulate a scenario where a few high-ranked trains are given a high delay. In this way, we can use our simulation to estimate the number of new train services to add to reduce delay propagation through sharing rolling stock.

We find that adding very few train services can significantly mitigate the delay in transmission. Specifically, the optimization technique bends the tail of the delays' power-law distribution, which is the most crucial part to reduce and control. Moreover, the optimization results using the agent-based simulation and the network approach are close in the procedure's first steps (the most significant). 

\section*{Text S2: Reduction over another line}
\begin{figure}[ht]
\centering
\includegraphics[width=14cm]{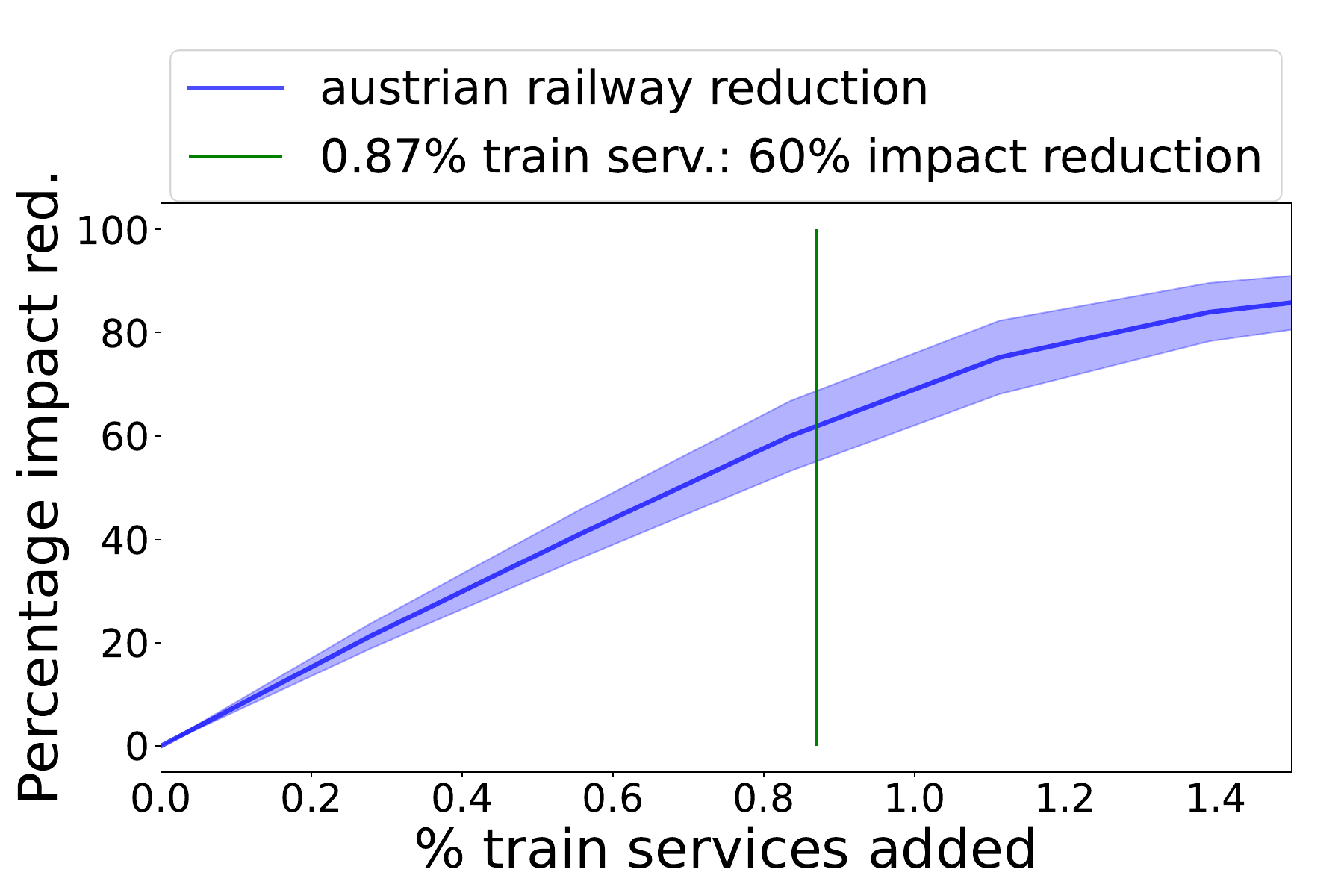}
\caption{\textbf{Optimisation line \ 10102} Network optimisation curve for line\ 10102 Wien-Salzburg over 30 random days in 2019. 
The blue line represents the average and standard deviation of the relative impact reduction as a result of adding new train services.
The vertical green line shows that adding the same percentage of train services with respect to our main result in Fig.~4 would reduce the impact by 60 \%.
The number of trains added would be $3$ over a total number of $380$.
This plot highlights that the optimization is more efficient on less busy lines.
\label{fig:10102}}
\end{figure}

In Fig.~\ref{fig:10102}, we show the NDP optimization procedure for line 10102 Wien-Salzburg over 30 random days. This plot shows that adding three rolling stocks would reduce the total impact by 60 \%. This suggests that, for lines with less traffic, removing links in the rolling stock layer for the most impactful trains reduces the total impact even more.
The curve reaches 100 \% when all the links are removed.

\section*{Text S3: Reduction over the whole national network}
\begin{figure}[ht]
\centering
\includegraphics[width=14cm]{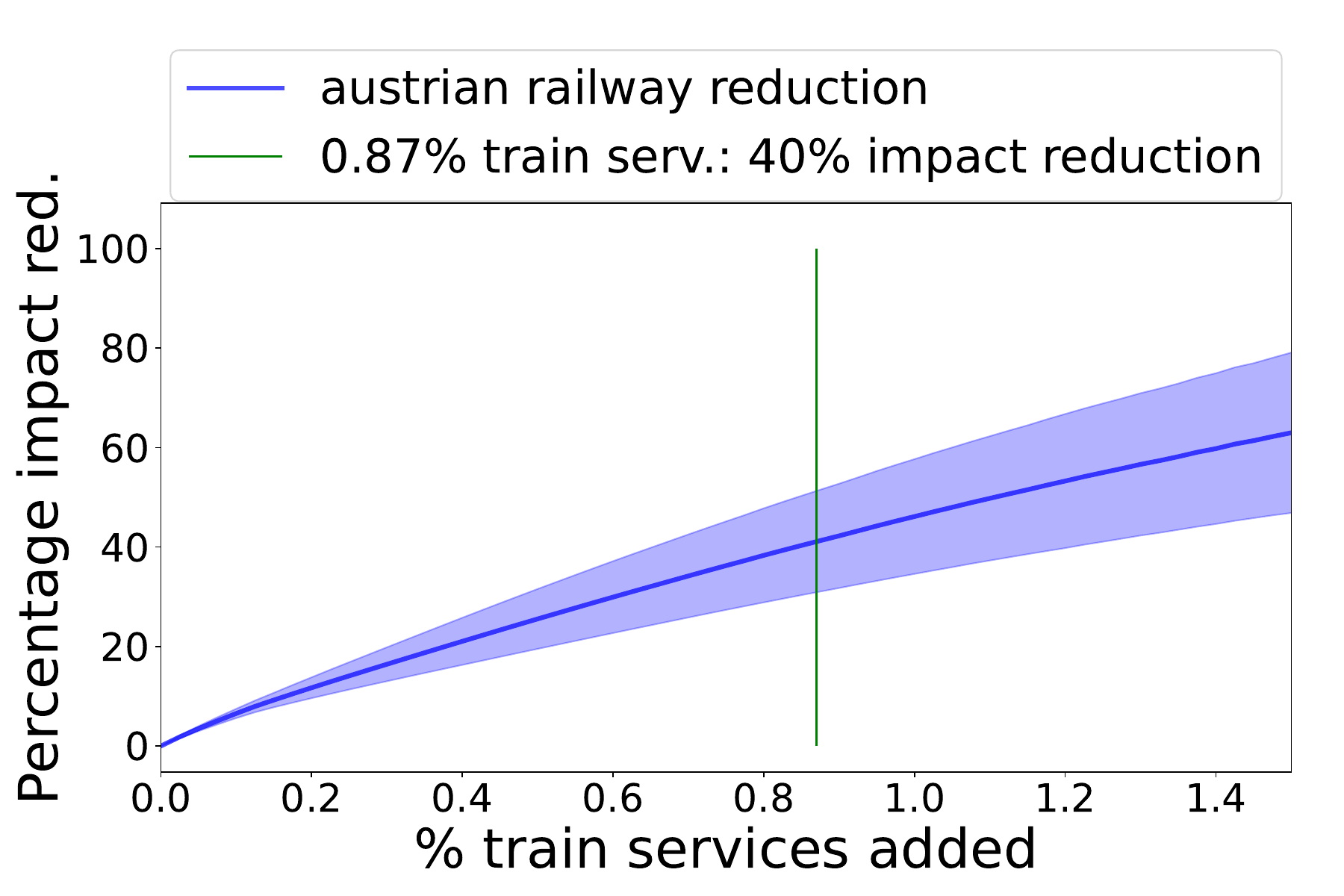}
\caption{\textbf{National Railway Optimisation}
    Optimization over the whole Nation using the Network technique.
    The blue line represents the average and standard deviation of the relative impact reduction as a result of adding new train services.
    The vertical green line shows that adding the same proportion of trains as in line Wr.Neustat-Wien HBF, the average gain is around \ 40 \%.
    The number of trains added would be $37$ over a total number of $4300$.
\label{fig:austria_reduction}}
\end{figure}

In Fig.~\ref{fig:austria_reduction} we show the NDP optimization procedure for the whole Austrian railway network for 30 random working days. 

This plot shows that adding three rolling stocks would reduce the total impact by 40 \%, adding 37 trains in total. This suggests that in most of the lines, removing three rolling stock contacts would improve the line even more than the main result of 20 \%, being that we reached that value for a line with a lot of traffic.
The curve reaches 100 \% when all the links are removed.

\subsection*{SI Text S4: Glossary of railway terms \label{si:glossary}}\hfill\\
In the following, we list some of the railway technical terms used in this paper alphabetically.
\begin{description}
    \item[block] \hfill \\
        A length of track of defined limits, the use of which by a movement is governed by block signals. In Austria, there may not be two physical trains on the same track of a block
    \item[buffer] \hfill \\
        a scheduled amount of time a train is allowed to wait at the station; It is frequently used in rail mobility to absorb part of accumulated delays
    \item[circulation] \hfill \\
        the usage of the rolling stock of another train service; Usually, different train services share the same physical rolling stock during the day; For example, given two last stop stations S and T, the same rolling stock can travel back and forth multiple times during the day, changing train service ID at each turnaround
    \item[conductor] \hfill \\
        a crew member responsible for operational and safety duties that do not involve the actual operation of the train
    \item[driver] \hfill \\
        a crew member responsible for operating train traction units
    \item[endogenous delay (primary delay)] \hfill \\
        the delay caused by trains that restrict resources to other trains. For example: if a train is on a block, the following train may not enter it; if the rolling stock of train service is shared with another, this latter has to wait for the arrival of the first one; if the personnel of a train is shared, the following train has to wait for it
    \item[exogenous delay (secondary delay)] \hfill \\
        a delay caused by factors external to the railway system. For example, bad weather conditions, passenger overcrowding, strikes, accidents, etc.
    \item[inspector] \hfill \\
        a crew member who is in charge of ticket control
    \item[infrastructure] \hfill \\
        the set of facilities that allow the transit of trains. It includes tracks, switches, platforms, sidings, and signals
    \item[line] \hfill \\
        the set of blocks with a common direction. Usually, it extends between two large cities
    \item[operational control points] \hfill \\
        special points scattered on the railway system allowing for train control. They usually host signals and are located at the boundary of blocks. Stations are also OCP
    \item[personnel] \hfill \\
        the set of people required for the operation of train services -- it includes drivers, conductors, and inspectors
    \item[platform] \hfill \\
        the area in stations where passengers may get on and off trains
    \item[primary delay] \hfill \\
        same as exogenous delay
    \item[rolling stock] \hfill \\
        the set of wheeled vehicles suitable for rail transport. It includes traction units, passenger wagons, and freight wagons
    \item[secondary delay] \hfill \\
        same as endogenous delay
    \item[section of line] \hfill \\
        same as a block
    \item[siding] \hfill \\
        track in a station that is used for maneuvering or parking trains. Sidings are often used  by freight trains. Some sidings may not allow for the transit of passenger trains for security reasons
    \item[signal] \hfill \\
        a device designed to control the transit of trains on blocks. It can be considered a kind of traffic light that enables transit to the next block when it is free
    \item[station] \hfill \\
        a station is an operational point where passengers may get on and off trains
    \item[track] \hfill \\
        a railway track is a set of two parallel rows of long pieces of steel on which trains can travel by resting their wheels.
        Some tracks are usually used for one direction only, and some others allow the transit of trains in both directions
    \item[traction unit] \hfill \\
        a vehicle responsible for the traction force of trains. It may be a locomotive with no passengers allowed or a railcar
    \item[train service] \hfill \\
        train services are the collection of services and resources needed for the train to run, e.g., rolling stock, infrastructure, and personnel. In the daily public schedule, train services are indicated by a unique identifier, e.g., ``RJ 123'' (train class: Rail Jet; train number: 123)
    \item[wagon] \hfill \\
        a vehicle for passenger transport.
\end{description}

\section*{Text S5: Differences in the two layers}
\begin{figure}[ht!]
\centering
\includegraphics[width=0.49\textwidth]{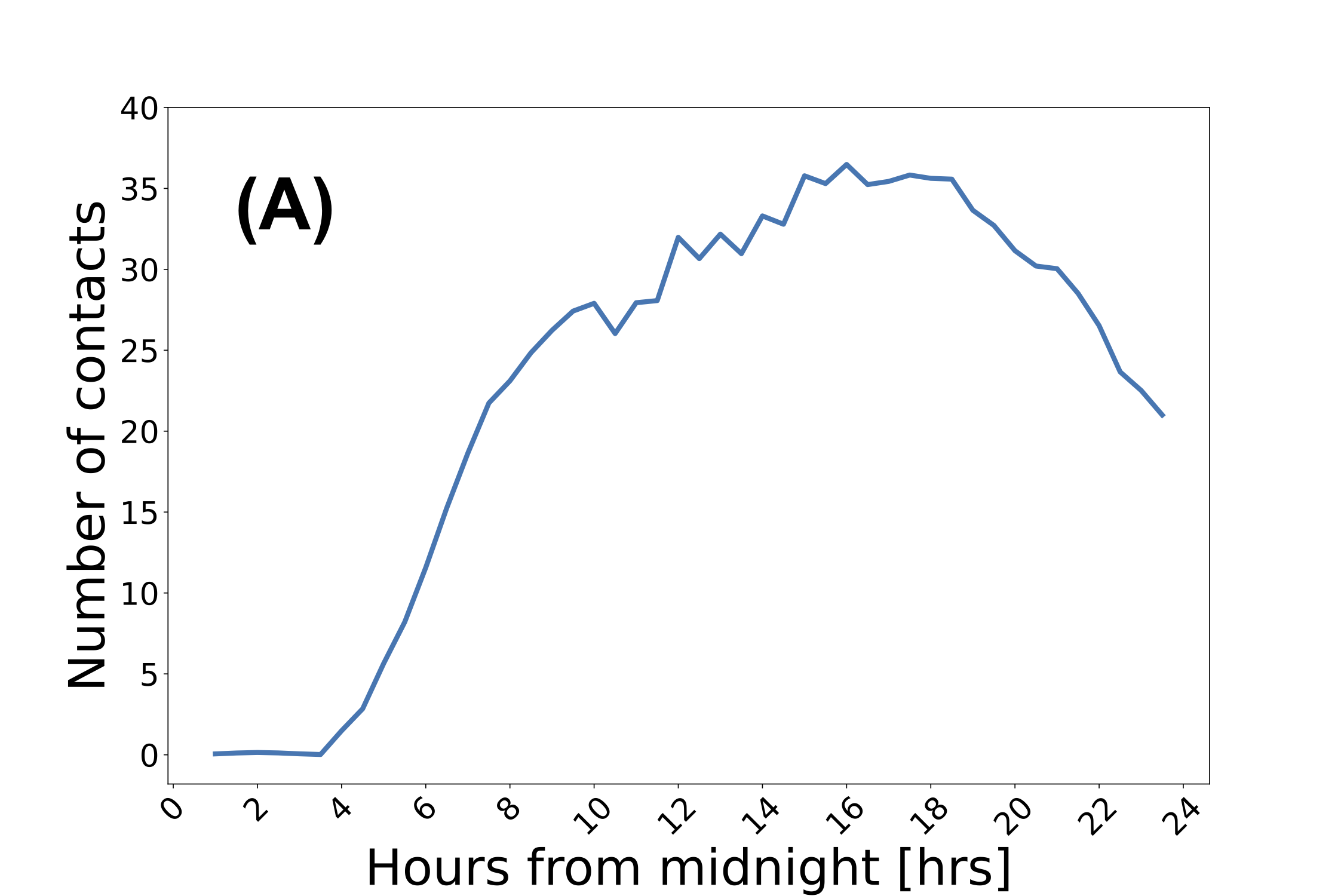}
\includegraphics[width=0.49\textwidth]{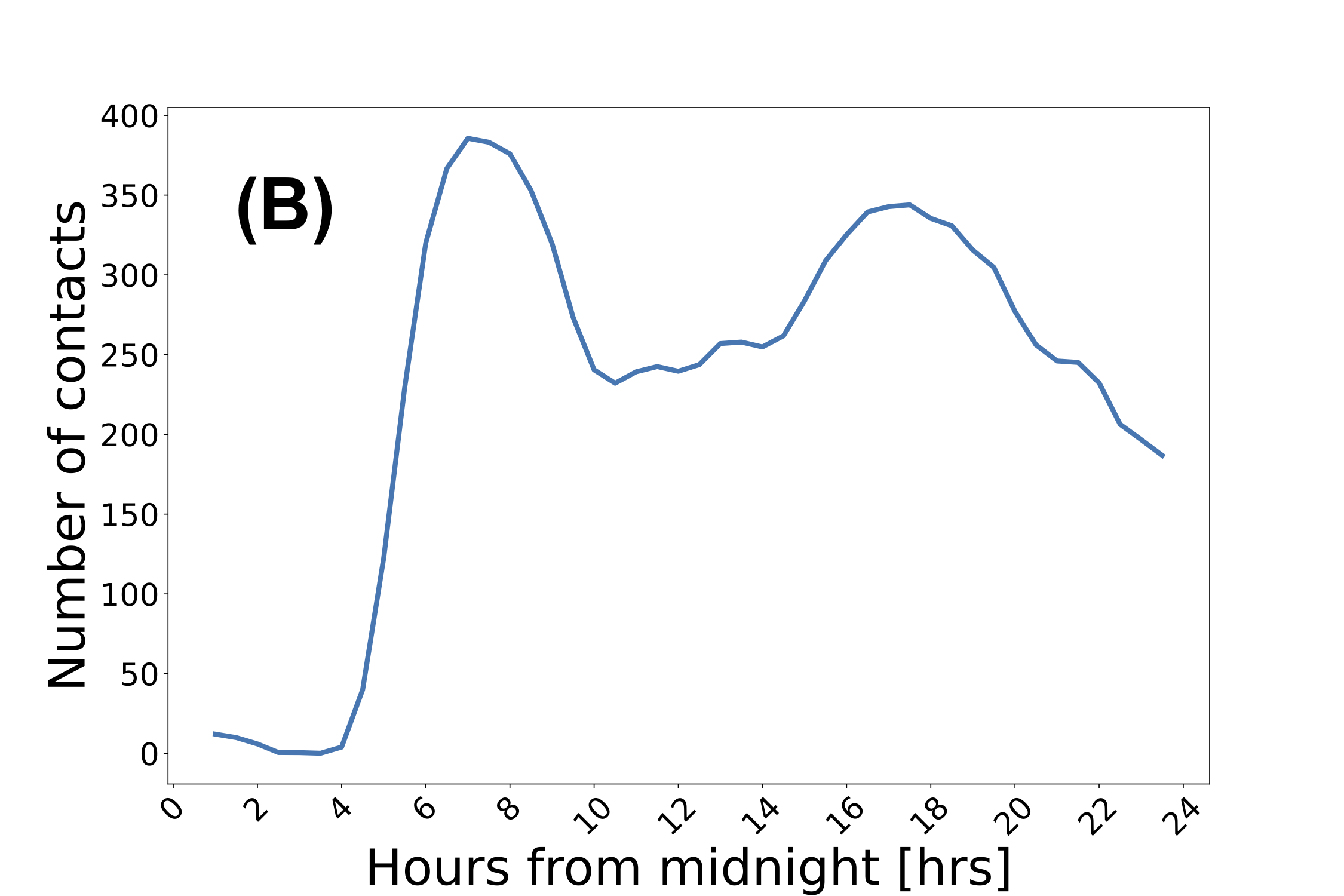}
\caption{ 
    \textbf{Density of contacts.}
    (A):
        The density of rolling stock contacts, i.e., the number of shared rolling stocks per hour.
    (B):
        The density of infrastructure contacts, i.e., the number of possible delay transmissions due to the shared infrastructure.
        The two peaks correspond to the morning and evening rush hours.
    The plots are averaged throughout the whole year and for working days only. The standard deviation is negligible.
\label{fig:densities}}
\end{figure}

All the results are based on a full year (2019) of working days.
In Fig.~\ref{fig:densities} we show the density of the rolling stock (A) and infrastructure layers (B).
We can observe that the amount of infrastructure contact is approximately one order of magnitude more significant.
Moreover, they reach their local maxima at different moments throughout a working day: infrastructure contacts follow the delay in peak hours.
This also reflects that the number of trains running in that moment of the day is higher.
Instead, the number of rolling stock contacts reaches a global maximum around 3 pm.

\begin{figure}[ht]
\centering
\includegraphics[width=14cm]{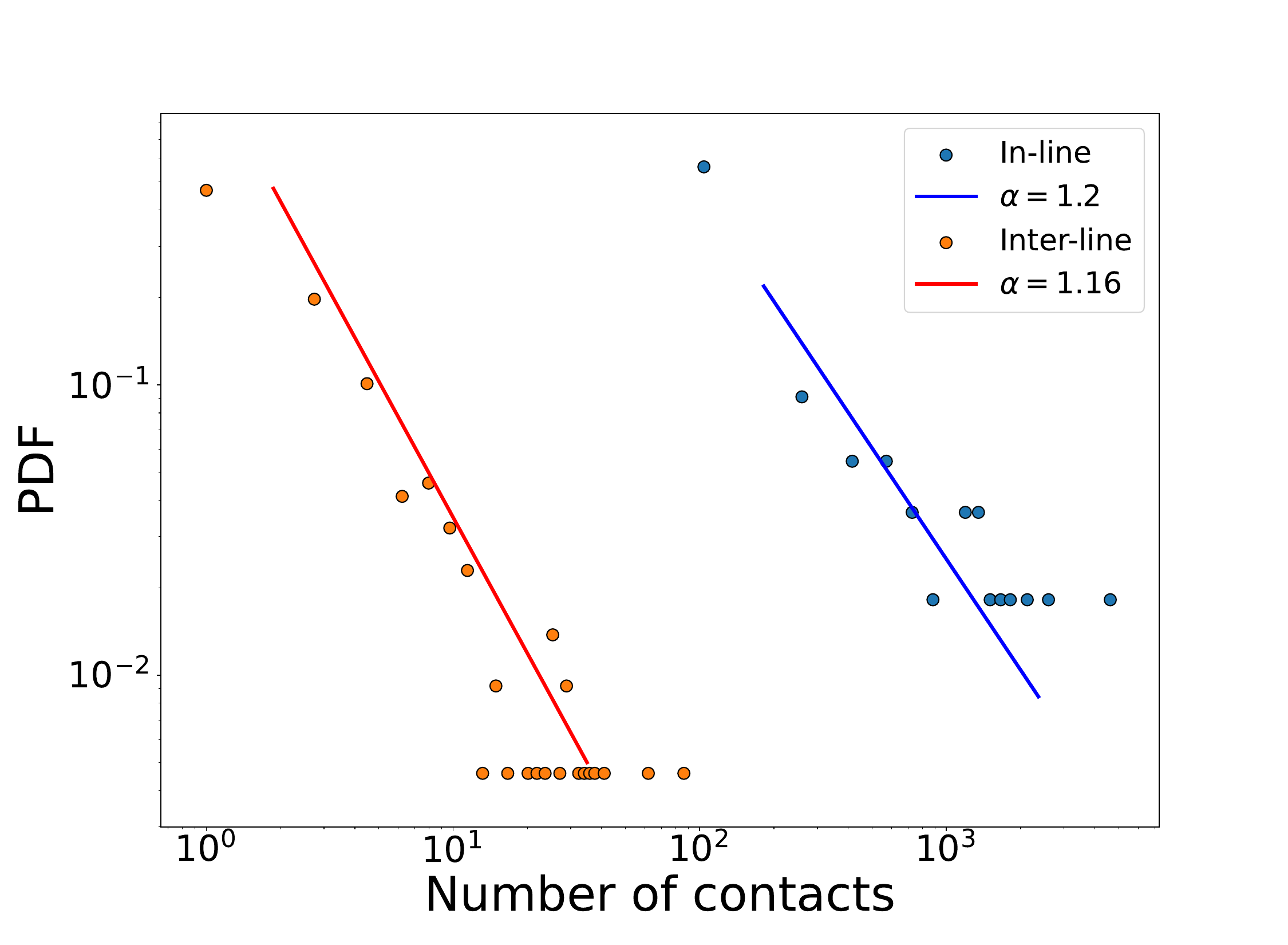}
\caption{\textbf{Distribution of contacts.}
        Distribution of contacts inside each line (blue) and between lines(orange). The PDF of the two show fat-tailed statistics. We fitted power-law distributions using the maximum entropy method, and report the exponents here.
        We observe that the two distributions are orders of magnitude different.
\label{fig:lines}}
\end{figure}

In Fig.~\ref{fig:lines}, we show the distribution of the number of contacts inside and between lines. 
This figure shows that the number of contacts inside each line is orders of magnitude higher than the contacts between lines.
This result, coupled with the preceding ones, suggests that compartmentalizing the national railway system and implementing optimized train services can greatly improve its performance, reduce delays, and minimize associated costs.

\begin{figure}[ht]
\centering
\includegraphics[width=14cm]{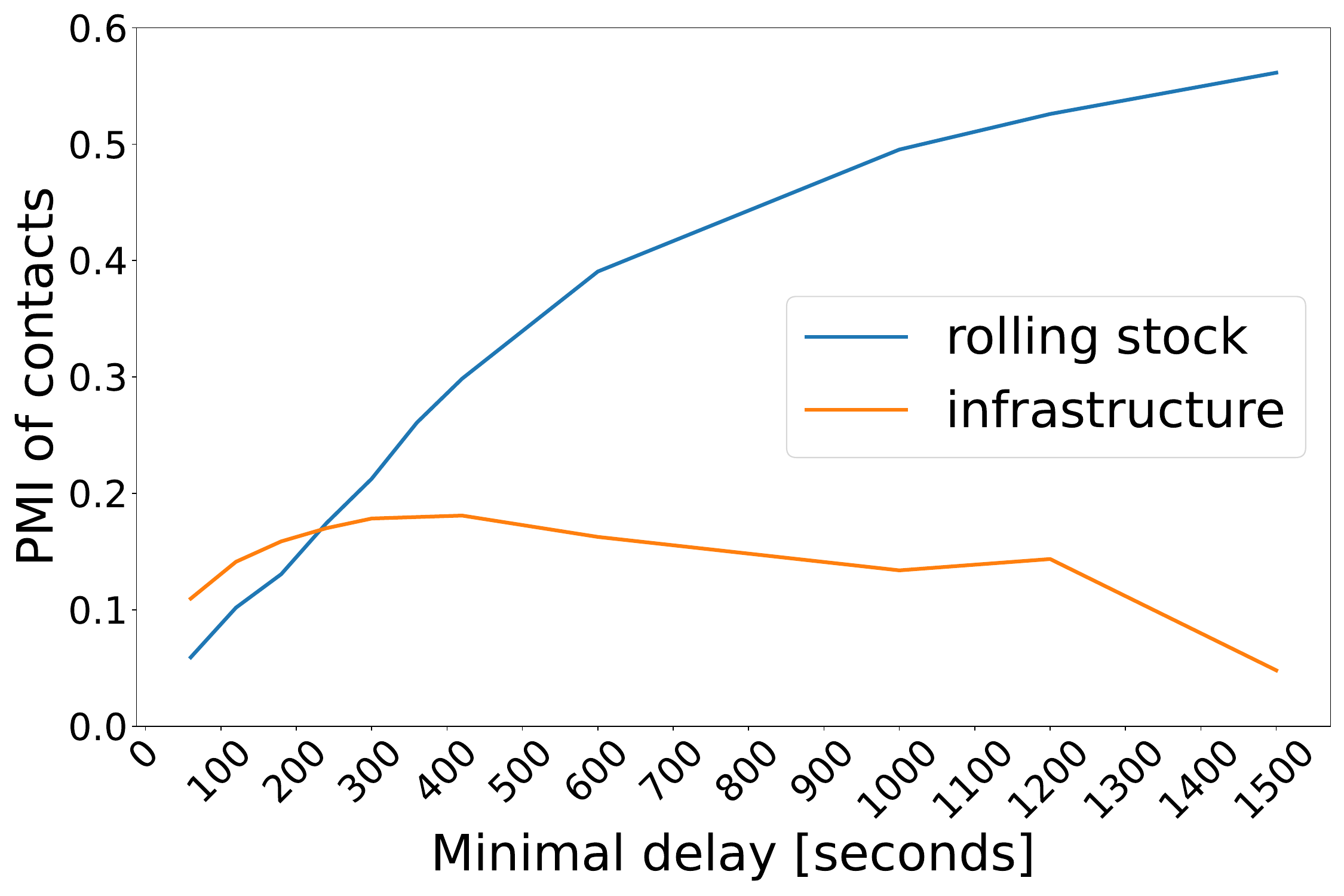}
\caption{\textbf{Pointwise Mutual Information of the Contacts.}
Pointwise Mutual Information of the Contact Events: for each point in the line plot, we consider the events in which the intensity of the delay is greater than a minimal amount, called \textit{minimal delay}.
}\label{fig:pmi}
\end{figure}

\begin{figure}[ht]
\centering
\includegraphics[width=14cm]{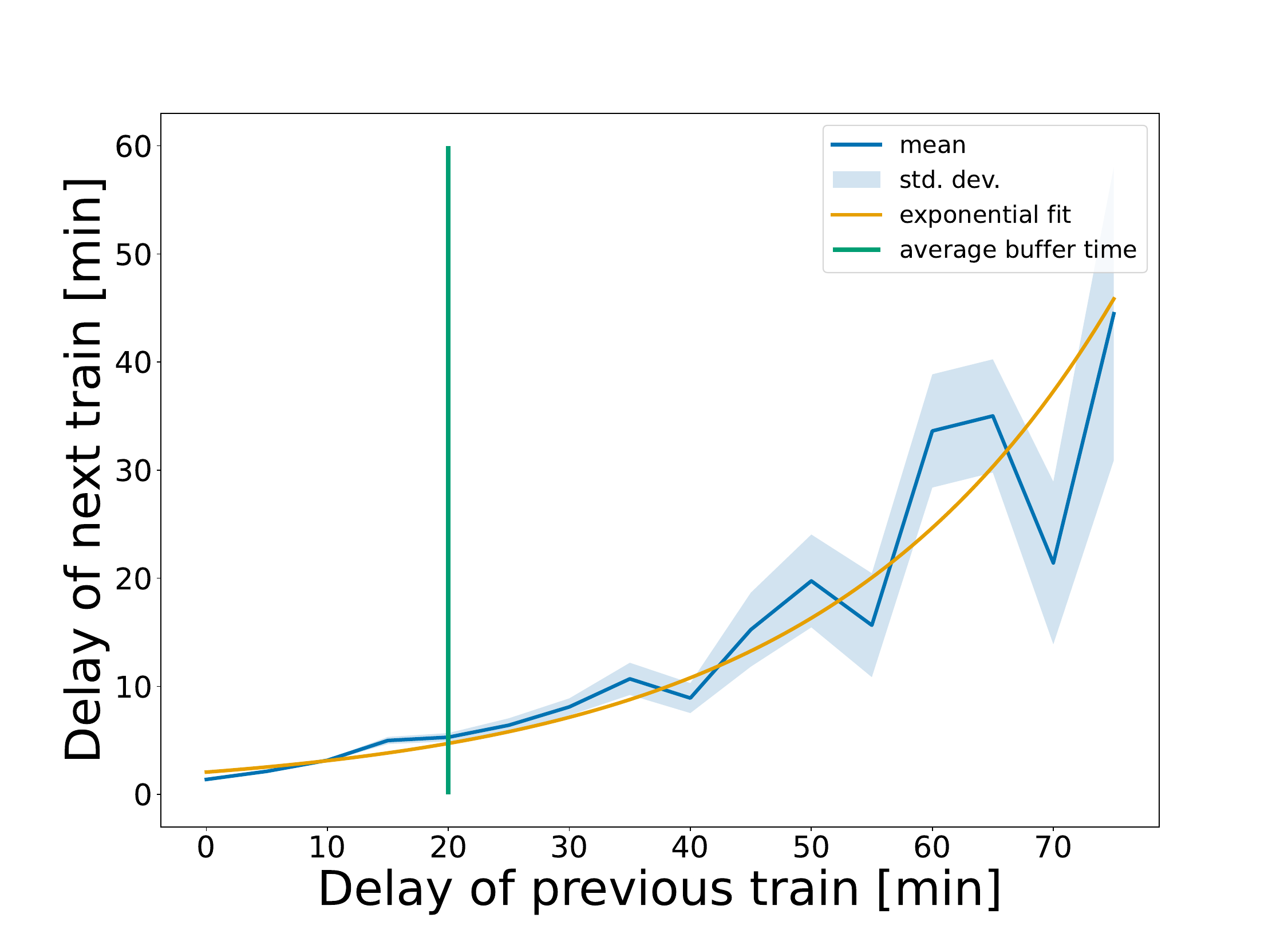}
\caption{\textbf{Delay influence in Rolling Stock Contacts.}Delay of the influencer train (x-axis) plotted with reference to the delay of the influenced train (y-axis). We see a rise in the dependency between the two, starting approximately after the average buffer time of 20 minutes (green vertical line). We plot the fit of an exponential function to the curve (yellow line).}\label{fig:binned}
\end{figure}

In Fig.~\ref{fig:binned}, we display the relation between the delay of the influencer train (on the x-axis) and the delay of the influenced train (y-axis).
We observe that as the influencer delay becomes more severe, also the delay of the influenced train becomes more relevant, fitting an exponential distribution.
This suggests that, as the influencer train delay grows, its influence on the influenced train overcomes mechanisms to cope with delays, for example, the buffer at the station.
It is impossible to see this relation for the infrastructure contacts, as shown in Fig.~\ref{fig:pmi}.

This picture displays the pointwise mutual information(PMI)~\cite{church1990word} between having the influencer train with a delay of at least a \emph{minimal delay} and the influenced train having a delay.
On the x-axis is displayed this minimal value; instead, on the y-axis is plotted the PMI.
Having a null PMI means that the two events are independent.
We can observe that for a low minimal delay, none of the contacts contain much information about the influenced train having a delay.
Instead, as the minimal delay grows, the rolling stock gains importance and the infrastructure doesn't.
This suggests that rolling stock contacts are more important and influential in spreading delays in railway networks.
It is impossible to neglect the infrastructure contacts due to their number and the fact that they generate the giant component of contacts.

\section*{Text S6: Ranking trains' impact using the ABS}

\begin{figure}[ht!]
\centering
\includegraphics[width=14cm]{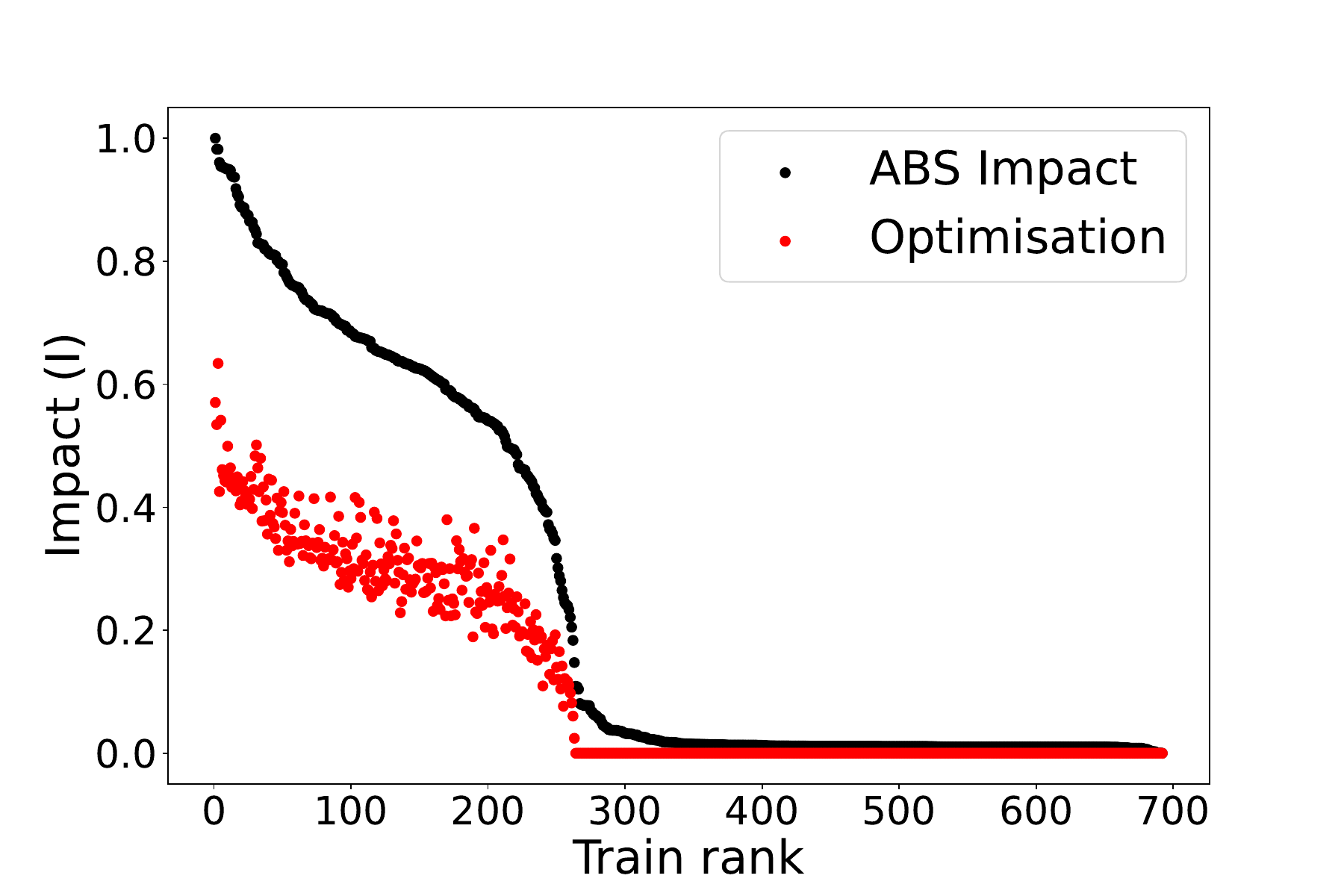}
\caption{\textbf{ABS rankplot of trains impact}
        Trains are ranked on the horizontal axis according to their simulation impact at the end of the day, calculated as the total delay of all trains at the end of their trajectory. 
        The blue line represents the results of the Agent-Based simulation considering both the layers of infrastructure and rolling stock. 
        The orange points consider the infrastructure network only and
        are ordered according to the rank defined by the blue line.
\label{fig:abs_rank}}
\end{figure}

Fig.~\ref{fig:abs_rank} complements Fig.4 in the main text showing the rankplot distribution of trains' impact computed by the ABS.
The procedure works as follows: we initialize the system in a way it simulates a normal working day, with no particular high delays.
Then, for each train, we artificially introduce a significant delay, corresponding to one hour in the figure, and compute the total delay of all the trains
at the end of the day.
As a result, we observe that some trains are significantly more impactful with respect to others, as shown by the black line.
Running the same procedure without the rolling stock contact layer, we obtain the red dots in the figure.
We observe a substantial reduction of the impact of the top-ranked trains by a factor of two.
However, we observe that the impact of trains after rank $\approx 270$, is approximately zero.
Instead, for our NDP it takes double the steps to reach non-influencer trains.
We believe this is due to the fact that for NDP we are diffusing the impact through the edges, meaning that each contact of the same layer is equally likely to spread impact with that intensity.
This makes more trains to be influencers.
Instead, in the ABS, handling each conflict event, some trains never spread the delay to others.

\section*{Text S7: NDP optimisation vs ABS optimisation}

\begin{figure}[ht]
\centering
\includegraphics[width=14cm]{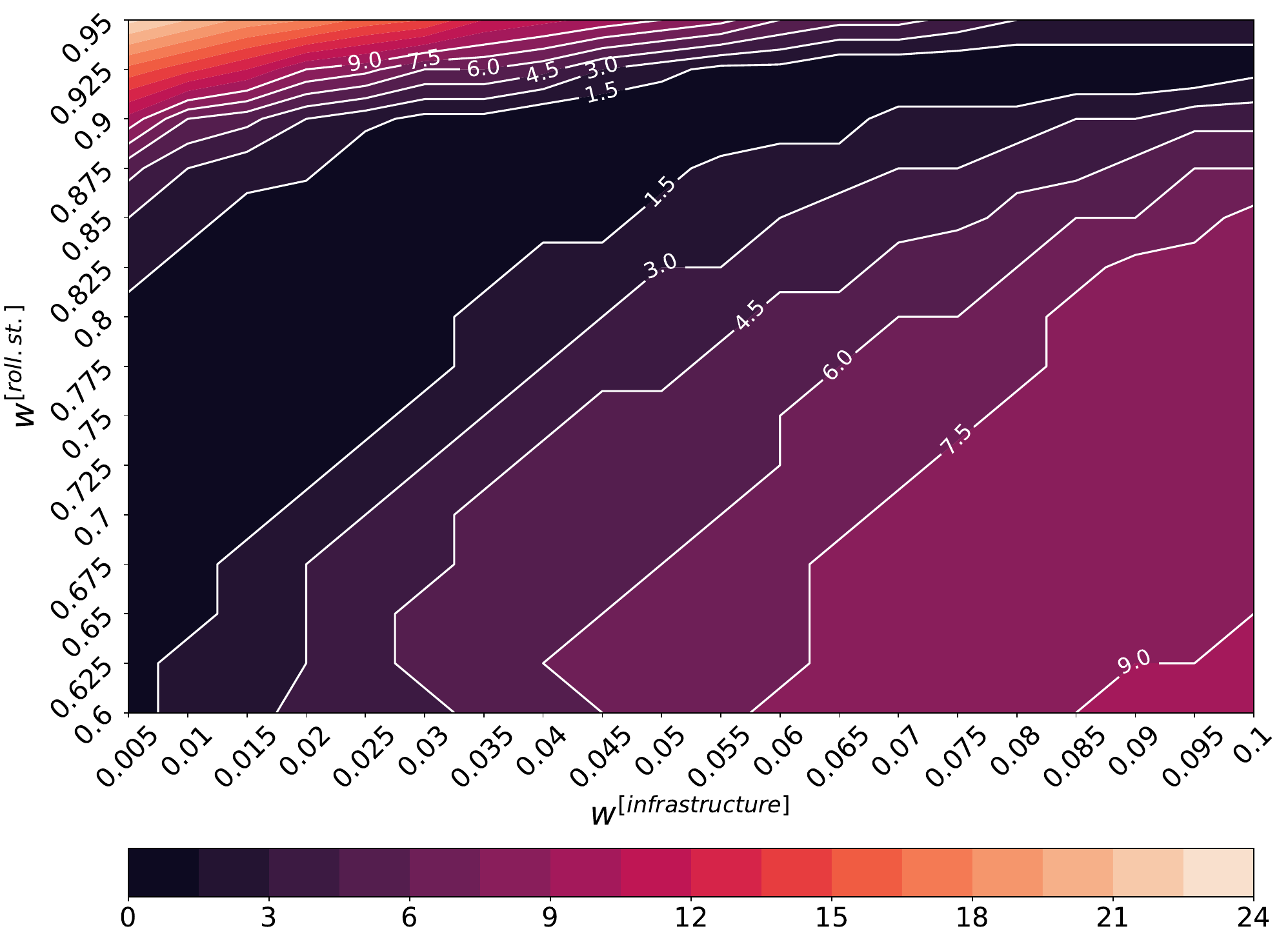}
\caption{ 
\textbf{Diffusion simulation approximating ABS results.}
    Contour plot representing the percentage of error between our NDP and the ABS on adding three train services.
    Each diffusion simulation depends on two parameters, namely the two diffusion coefficients over the two layers ($w^{[roll.st.]}$ and $w^{[infrastructure]}$).
    Each value is the result of the absolute difference between the means of the two optimization techniques when $3$ train services are added.
    The results are shown for a trimester (from January 1$^{st}$ to March 31$^{st}$ 2019) in the Austrian S\"udbahn railway line from Vienna Central Station to Wiener Neustadt.
\label{fig:phase}}
\end{figure}

In Fig.~\ref{fig:phase}, we show the percentage of error between the NDP and ABS impact reduction procedures.
The picture shows that there is a wide band of combinations of the two diffusion parameters (one for each layer) regulating the NDP process in which the error is around \ 1 \%.
These values show that the value of the diffusion parameter over the rolling stock layer needs to be much higher than the one over the infrastructure layer in order to reproduce the same reduction of the ABS.
This result is in line with the previous plots.

\begin{thebibliography}{10}

\bibitem{de2011world}
Luca De~Benedictis and Lucia Tajoli.
\newblock The world trade network.
\newblock {\em The World Economy}, 34(8):1417--1454, 2011.

\bibitem{boss2004network}
Michael Boss, Helmut Elsinger, Martin Summer, and Stefan Thurner~4.
\newblock Network topology of the interbank market.
\newblock {\em Quantitative finance}, 4(6):677--684, 2004.

\bibitem{thurner2013debtrank}
Stefan Thurner and Sebastian Poledna.
\newblock Debtrank-transparency: Controlling systemic risk in financial
  networks.
\newblock {\em Scientific reports}, 3(1):1888, 2013.

\bibitem{Harland2001ATO}
Christine~Mary Harland, Richard Lamming, Jurong Zheng, and Thomas Johnsen.
\newblock A taxonomy of supply networks.
\newblock {\em Journal of Supply Chain Management}, 37:21--27, 2001.

\bibitem{glassman2011geo}
Jim Glassman.
\newblock The geo-political economy of global production networks.
\newblock {\em Geography Compass}, 5(4):154--164, 2011.

\bibitem{alessandretti2020scales}
Laura Alessandretti, Ulf Aslak, and Sune Lehmann.
\newblock The scales of human mobility.
\newblock {\em Nature}, 587(7834):402--407, 2020.

\bibitem{battiston2016price}
Stefano Battiston, Guido Caldarelli, Robert~M May, Tarik Roukny, and Joseph~E
  Stiglitz.
\newblock The price of complexity in financial networks.
\newblock {\em Proceedings of the National Academy of Sciences},
  113(36):10031--10036, 2016.

\bibitem{al2016economic}
Amro Al~Kazimi and Cameron~A Mackenzie.
\newblock The economic costs of natural disasters, terrorist attacks, and other
  calamities: An analysis of economic models that quantify the losses caused by
  disruptions.
\newblock In {\em 2016 IEEE Systems and Information Engineering Design
  Symposium (SIEDS)}, pages 32--37. IEEE, 2016.

\bibitem{tsuchiya2007economic}
Satoshi Tsuchiya, Hirokazu Tatano, and Norio Okada.
\newblock Economic loss assessment due to railroad and highway disruptions.
\newblock {\em Economic Systems Research}, 19(2):147--162, 2007.

\bibitem{boss2004contagion}
Michael Boss, Martin Summer, and Stefan Thurner.
\newblock Contagion flow through banking networks.
\newblock In {\em Computational Science-ICCS 2004: 4th International
  Conference, Krak{\'o}w, Poland, June 6-9, 2004, Proceedings, Part III 4},
  pages 1070--1077. Springer, 2004.

\bibitem{Hirshleifer2003HerdBA}
David Hirshleifer and Siew Hong~Teoh.
\newblock Herd behaviour and cascading in capital markets: a review and
  synthesis.
\newblock {\em European Financial Management}, 9(1):25--66, 2003.

\bibitem{Crucitti2003ModelFC}
Paolo Crucitti, Vito Latora, and Massimo Marchiori.
\newblock Model for cascading failures in complex networks.
\newblock {\em Physical review. E, Statistical, nonlinear, and soft matter
  physics}, 69 4 Pt 2:045104, 2003.

\bibitem{BorgeHolthoefer2013CascadingBI}
Javier Borge-Holthoefer, Raquel~Alvarez Ba{\~n}os, Sandra
  Gonz{\'a}lez-Bail{\'o}n, and Yamir Moreno.
\newblock Cascading behaviour in complex socio-technical networks.
\newblock {\em J. Complex Networks}, 1:3--24, 2013.

\bibitem{thurner2018introduction}
Stefan Thurner, Rudolf Hanel, and Peter Klimek.
\newblock {\em Introduction to the theory of complex systems}.
\newblock Oxford University Press, 2018.

\bibitem{briggs2007modelling}
Keith Briggs and Christian Beck.
\newblock Modelling train delays with q-exponential functions.
\newblock {\em Physica A: Statistical Mechanics and its Applications},
  378(2):498--504, 2007.

\bibitem{monechiComplexDelayDynamics2018}
Bernardo Monechi, Pietro Gravino, Riccardo Di~Clemente, and Vito~D.~P.
  Servedio.
\newblock Complex delay dynamics on railway networks from universal laws to
  realistic modelling.
\newblock {\em EPJ Data Science}, 7(1):35, December 2018.

\bibitem{havlin2014vulnerability}
S~Havlin, DY~Kenett, A~Bashan, J~Gao, and HE~Stanley.
\newblock Vulnerability of network of networks.
\newblock {\em The European Physical Journal Special Topics}, 223:2087--2106,
  2014.

\bibitem{dekker2021cascading}
Mark~M Dekker and Debabrata Panja.
\newblock Cascading dominates large-scale disruptions in transport over complex
  networks.
\newblock {\em PLoS One}, 16(1):e0246077, 2021.

\bibitem{bhatia2015network}
Udit Bhatia, Devashish Kumar, Evan Kodra, and Auroop~R Ganguly.
\newblock Network science based quantification of resilience demonstrated on
  the indian railways network.
\newblock {\em PloS one}, 10(11):e0141890, 2015.

\bibitem{schipperDifferencesSimilaritiesEuropean2018}
Danny Schipper and Lasse Gerrits.
\newblock Differences and similarities in {{European}} railway disruption
  management practices.
\newblock {\em Journal of Rail Transport Planning \& Management}, 8(1):42--55,
  June 2018.

\bibitem{buldyrev2010catastrophic}
Sergey~V Buldyrev, Roni Parshani, Gerald Paul, H~Eugene Stanley, and Shlomo
  Havlin.
\newblock Catastrophic cascade of failures in interdependent networks.
\newblock {\em Nature}, 464(7291):1025--1028, 2010.

\bibitem{ball2016two}
Robin Ball, Debabrata Panja, and Gerard~T Barkema.
\newblock A two component railway model exhibiting service collapse.
\newblock {\em ""}, "", 2016.

\bibitem{leobons2019assessing}
Camila~Maestrelli Leobons, V{\^a}nia Barcellos~Gouv{\^e}a Campos, and
  Renata~Albergaria de~Mello~Bandeira.
\newblock Assessing urban transportation systems resilience: a proposal of
  indicators.
\newblock {\em Transportation research procedia}, 37:322--329, 2019.

\bibitem{lordan2015robustness}
Oriol Lordan, Jose~M Sallan, Pep Simo, and David Gonzalez-Prieto.
\newblock Robustness of airline alliance route networks.
\newblock {\em Communications in Nonlinear Science and Numerical Simulation},
  22(1-3):587--595, 2015.

\bibitem{pagani2019resilience}
Alessio Pagani, Guillem Mosquera, Aseel Alturki, Samuel Johnson, Stephen
  Jarvis, Alan Wilson, Weisi Guo, and Liz Varga.
\newblock Resilience or robustness: identifying topological vulnerabilities in
  rail networks.
\newblock {\em Royal Society open science}, 6(2):181301, 2019.

\bibitem{sun2019exploring}
Qipeng Sun, Xiaozhuang Guo, Wenjing Jiang, Haiying Ding, Tingzhen Li, and
  Xingbo Xu.
\newblock Exploring the node importance and its influencing factors in the
  railway freight transportation network in china.
\newblock {\em Journal of Advanced Transportation}, 2019, 2019.

\bibitem{wang2013vulnerability}
Junjie Wang, Yishuai Li, Jingyu Liu, Kun He, and Pu~Wang.
\newblock Vulnerability analysis and passenger source prediction in urban rail
  transit networks.
\newblock {\em PloS one}, 8(11):e80178, 2013.

\bibitem{kecman2015stochastic}
Pavle Kecman, Francesco Corman, Anders Peterson, and Martin Joborn.
\newblock Stochastic prediction of train delays in real-time using bayesian
  networks.
\newblock In {\em Conference on Advanced Systems in Public Transport (CASPT
  2015)}. CASPT, 2015.

\bibitem{dekker2022modelling}
Mark~M Dekker, Alexey~N Medvedev, Jan Rombouts, Grzegorz Siudem, and Liubov
  Tupikina.
\newblock Modelling railway delay propagation as diffusion-like spreading.
\newblock {\em EPJ Data Science}, 11(1):44, 2022.

\bibitem{dekkerPredictingTransitionsMacroscopic2019}
Mark~M. Dekker, Debabrata Panja, Henk~A. Dijkstra, and Stefan~C. Dekker.
\newblock Predicting transitions across macroscopic states for railway systems.
\newblock {\em PLOS ONE}, 14(6):e0217710, June 2019.

\bibitem{onetoTrainDelayPrediction2018}
Luca Oneto, Emanuele Fumeo, Giorgio Clerico, Renzo Canepa, Federico Papa, Carlo
  Dambra, Nadia Mazzino, and Davide Anguita.
\newblock Train {{Delay Prediction Systems}}: {{A Big Data Analytics
  Perspective}}.
\newblock {\em Big Data Research}, 11:54--64, March 2018.

\bibitem{fleurquin2013systemic}
Pablo Fleurquin, Jos{\'e}~J Ramasco, and Victor~M Eguiluz.
\newblock Systemic delay propagation in the us airport network.
\newblock {\em Scientific reports}, 3(1):1159, 2013.

\bibitem{ludvigsen2014extreme}
Johanna Ludvigsen and Ronny Kl{\ae}boe.
\newblock Extreme weather impacts on freight railways in europe.
\newblock {\em Natural hazards}, 70:767--787, 2014.

\bibitem{goverdeDelayPropagationAlgorithm2010}
Rob~M.P. Goverde.
\newblock A delay propagation algorithm for large-scale railway traffic
  networks.
\newblock {\em Transportation Research Part C: Emerging Technologies},
  18(3):269--287, June 2010.

\bibitem{gambardellaAgentbasedPlanningSimulation2002}
Luca~Maria Gambardella, Andrea~E. Rizzoli, and Petra Funk.
\newblock Agent-based {{Planning}} and {{Simulation}} of {{Combined
  Rail}}/{{Road Transport}}.
\newblock {\em SIMULATION}, 78(5):293--303, May 2002.

\bibitem{monechiCongestionTransitionAir2015}
Bernardo Monechi, Vito D.~P. Servedio, and Vittorio Loreto.
\newblock Congestion {{Transition}} in {{Air Traffic Networks}}.
\newblock {\em PLOS ONE}, 10(5):e0125546, May 2015.

\bibitem{bukerStochasticModellingDelay2012}
Thorsten B{\"u}ker and Bernhard Seybold.
\newblock Stochastic modelling of delay propagation in large networks.
\newblock {\em Journal of Rail Transport Planning \& Management},
  2(1-2):34--50, November 2012.

\bibitem{meester2007stochastic}
Ludolf~E Meester and Sander Muns.
\newblock Stochastic delay propagation in railway networks and phase-type
  distributions.
\newblock {\em Transportation Research Part B: Methodological}, 41(2):218--230,
  2007.

\bibitem{pyrgiotis2013modelling}
Nikolas Pyrgiotis, Kerry~M Malone, and Amedeo Odoni.
\newblock Modelling delay propagation within an airport network.
\newblock {\em Transportation Research Part C: Emerging Technologies},
  27:60--75, 2013.

\bibitem{campanelli2014modeling}
B~Campanelli, P~Fleurquin, VM~Eguiluz, JJ~Ramasco, A~Arranz, I~Etxebarria, and
  C~Ciruelos.
\newblock Modeling reactionary delays in the european air transport network.
\newblock {\em Proceedings of the Fourth SESAR Innovation Days, Schaefer D
  (Ed.), Madrid}, 1, 2014.

\bibitem{wei2015modeling}
Dali Wei, Hongchao Liu, and Yong Qin.
\newblock Modeling cascade dynamics of railway networks under inclement
  weather.
\newblock {\em Transportation Research Part E: Logistics and Transportation
  Review}, 80:95--122, 2015.

\bibitem{Sen_Dasgupta_Chatterjee_Sreeram_Mukherjee_Manna_2003}
Parongama Sen, Subinay Dasgupta, Arnab Chatterjee, P.~A. Sreeram, G.~Mukherjee,
  and S.~S. Manna.
\newblock Small-world properties of the indian railway network.
\newblock {\em Physical Review E}, 67(3):036106, Mar 2003.

\bibitem{oneto2017dynamic}
Luca Oneto, Emanuele Fumeo, Giorgio Clerico, Renzo Canepa, Federico Papa, Carlo
  Dambra, Nadia Mazzino, and Davide Anguita.
\newblock Dynamic delay predictions for large-scale railway networks: Deep and
  shallow extreme learning machines tuned via thresholdout.
\newblock {\em IEEE Transactions on Systems, Man, and Cybernetics: Systems},
  47(10):2754--2767, 2017.

\bibitem{zinser2018comparison}
Markus Zinser, Torsten Betz, Jennifer Warg, Emma Solinen, and Markus Bohlin.
\newblock Comparison of microscopic and macroscopic approaches to simulating
  the effects of infrastructure disruptions on railway networks.
\newblock In {\em Transport Research Arena Conference}. Zenodo, 2018.

\bibitem{rossler2020simulation}
Matthias R{\"o}{\ss}ler, Matthias Wastian, Anna Jellen, Sarah Frisch, Dominic
  Weinberger, Philipp Hungerl{\"a}nder, Martin Bicher, and Niki Popper.
\newblock Simulation and optimization of traction unit circulations.
\newblock In {\em 2020 Winter Simulation Conference (WSC)}, pages 90--101.
  IEEE, 2020.

\bibitem{nash2004railroad}
Andrew Nash and Daniel Huerlimann.
\newblock Railroad simulation using opentrack.
\newblock {\em WIT Transactions on The Built Environment}, 74, 2004.

\bibitem{johansson2022microscopic}
Ingrid Johansson, Carl-William Palmqvist, Hans Sipil{\"a}, Jennifer Warg, and
  Markus Bohlin.
\newblock Microscopic and macroscopic simulation of early freight train
  departures.
\newblock {\em Journal of Rail Transport Planning \& Management}, 21:100295,
  2022.

\bibitem{nakayama2018optically}
Yu~Nakayama, Kazuki Maruta, Takuya Tsutsumi, and Kaoru Sezaki.
\newblock Optically backhauled moving network for local trains: Architecture
  and scheduling.
\newblock {\em IEEE Access}, 6:31023--31036, 2018.

\bibitem{barbour2020enhanced}
William Barbour, Shankara Kuppa, and Daniel~B Work.
\newblock Enhanced data reconciliation of freight rail dispatch data.
\newblock {\em Journal of Rail Transport Planning \& Management}, 14:100193,
  2020.

\bibitem{Borgatti_2005}
Stephen~P. Borgatti.
\newblock Centrality and network flow.
\newblock {\em Social Networks}, 27(1):55–71, Jan 2005.

\bibitem{sajjadi2022structural}
Sina Sajjadi, Pourya~Toranj Simin, Mehrzad Shadmangohar, Basak Taraktas, Ulya
  Bayram, Maria~V. Ruiz-Blondet, and Fariba Karimi.
\newblock Structural inequalities exacerbate infection disparities: A
  computational approach, 2022.
\newblock eprint arXiv:2205.04361.

\bibitem{barrat2008dynamical}
Alain Barrat, Marc Barthelemy, and Alessandro Vespignani.
\newblock {\em Dynamical processes on complex networks}.
\newblock Cambridge university press, 2008.

\bibitem{goverdeRailwayTimetableStability2007}
Rob~M.P. Goverde.
\newblock Railway timetable stability analysis using max-plus system theory.
\newblock {\em Transportation Research Part B: Methodological}, 41(2):179--201,
  February 2007.

\end{thebibliography}

\end{document}